\documentclass[11pt]{article}
\pdfoutput=1 

\usepackage{jcappub} 

\usepackage{amssymb}
\usepackage{xcolor}
\usepackage[normalem]{ulem}

\usepackage[left=3cm,right=3cm,top=3cm,bottom=3cm]{geometry}
\usepackage[toc,page]{appendix}
\usepackage{feynmf}
\usepackage{commath}
\usepackage{tikz}
\usepackage{hyperref}
\usepackage{slashed}
\usetikzlibrary{decorations.pathmorphing,decorations.markings,plotmarks,quotes,angles}

\newcommand{\bea}{\begin{eqnarray}}
\newcommand{\eea}{\end{eqnarray}}

\title{
Constraints on cosmic-ray boosted dark matter with realistic cross section
}

\author[a]{Atanu Guha}
\author[a]{and Jong-Chul Park}

\affiliation[a]{Department of Physics and Institute of Quantum Systems, \protect\\ Chungnam National University, Daejeon, 34134, Republic of Korea}
\emailAdd{atanu@cnu.ac.kr}
\emailAdd{jcpark@cnu.ac.kr}

\abstract{Sub-MeV cold dark-matter particles are unable to produce electronic recoil in conventional dark-matter direct detection experiments such as XENONnT and LUX-ZEPLIN above the detector threshold. 
The mechanism of boosted dark matter comes into picture to constrain the parameter space of such low mass dark matter from direct detection experiments. 
We consider the effect of the leading components of cosmic rays to boost the cold dark matter, which results in significant improvements on the exclusion limits compared to the existing ones. 
To present concrete study results, we choose to work on models consisting of a dark-matter particle $\chi$ with an additional $U(1)'$ gauge symmetry including the secluded dark photon, $U(1)_{B-L}$, and $U(1)_{L_e-L_\mu}$.
We find that the energy dependence of the scattering cross section plays a crucial role in improving the constraints. 
In addition, we systematically estimate the Earth shielding effect on boosted dark matter in losing energy while traveling to the underground detector through the Earth. 
}

\begin{document}
\maketitle
\flushbottom

\section{Introduction}

 Various astrophysical evidences strongly support the notion of the existence of large amount of dark matter (DM) in our Universe~\cite{Planck:2015fie,Bertone:2004pz}. 
 Even though primary indication comes only from the gravitational effects, efforts have been made to extract some information about the interaction between DM particles and standard model (SM) particles. 
 One of the most promising avenues is to observe the recoil spectrum of SM targets when DM particles scatter off them~\cite{Goodman:1984dc}. 
 This idea has been implemented in the leading direct detection experiments~\cite{DAMA:2008jlt, CRESST:2015txj, DarkSide:2018ppu, SuperCDMS:2018mne, XENONCollaboration:2023orw, LUX:2022vee, PandaX-4T:2021bab}. 
 
 The most stringent constraint on the spin-independent elastic scattering cross section of cold DM with nucleon is now given as $\sim 5 \times 10^{-48}~\rm{cm^2}$ at 30 GeV of DM mass~\cite{LUX-ZEPLIN:2022xrq}. 
 The strongest limit on DM-electron elastic scattering cross section is given by $\sim 5 \times 10^{-41}~\rm{cm^2}$ at 200 MeV of DM mass~\cite{XENON:2022ltv}. 
 Direct detection experiments lose sensitivity for DM with low mass as energy transfer is less efficient for the lighter cold DM to overcome their detection threshold energies. 
 The lowest mass explored by direct detection experiments till date is $\sim0.5$ MeV when DM-electron interaction is considered~\cite{DAMIC-M:2023gxo, SuperCDMS:2020ymb, SENSEI:2020dpa, 
 EDELWEISS:2020fxc}, while $\sim0.2$ GeV when DM-nucleon spin-independent interaction is considered~\cite{CRESST:2019jnq}. 
 In our solar neighbourhood, the average velocity of halo DM is approximately $10^{-3}$ of the speed of light which is the primary cause for the low mass cold DM to fail to induce recoils above the detector threshold energy. 
 Quantitatively, the XENONnT experiment loses sensitivities for sub-GeV DM in case of nuclear recoils and for sub-MeV DM in case of electron recoils.
 
 To overcome this difficulty, various classes of boosted dark matter (BDM) have been proposed in literature.\footnote{Many new direct detection ideas with low detection threshold energy have been proposed targeting sub-MeV DM~\cite{Hochberg:2015pha, Schutz:2016tid, Hochberg:2017wce, Knapen:2017ekk, Hochberg:2019cyy, Kim:2020bwm}.} 
 Among them, two component DM models with a hierarchical mass spectrum~\cite{Belanger:2011ww, Agashe:2014yua, Berger:2014sqa, Kong:2014mia, Cherry:2015oca, Necib:2016aez, Alhazmi:2016qcs, Kim:2016zjx, Giudice:2017zke, Chatterjee:2018mej, Kim:2018veo, Aoki:2018gjf, Kim:2019had, Kim:2020ipj, DeRoeck:2020ntj, Alhazmi:2020fju}, charged cosmic-ray boosted DM (CRBDM)~\cite{Bringmann:2018cvk, Ema:2018bih, Cappiello:2019qsw, Dent:2019krz, Wang:2019jtk, Cao:2020bwd, Jho:2020sku, Cho:2020mnc, Dent:2020syp, Xia:2021vbz, Ghosh:2021vkt, Bardhan:2022bdg,Alvey:2022pad,Maity:2022exk,Bell:2023sdq}, and cosmic-ray neutrino BDM ($\nu$BDM)~\cite{Jho:2021rmn, Das:2021lcr, Chao:2021orr, Lin:2022dbl,DeRomeri:2023ytt} have received  special attention in recent times.
 Moreover, numerous phenomenological studies of BDM motivated several neutrino and DM experiments  to study and report dedicated search results of BDM~\cite{Super-Kamiokande:2017dch, COSINE-100:2018ged, PandaX-II:2021kai, CDEX:2022fig, Super-Kamiokande:2022ncz}.
 We consider the effect of the leading components of cosmic rays~\cite{Workman:2022ynf,PhysRevD.98.030001} to boost the cold DM, and find that there is significant improvements on the exclusion limits compared to the existing literature. 
 
 Working with the energy independent constant cross section for the whole mass range of DM provides a rough estimation about the exclusion region in the DM mass vs DM-SM interacting cross section parameter space~\cite{Ghosh:2021vkt}. 
 However, to present a concrete example, we choose to work on models described below consisting of a Dirac fermion $\chi$ with a new $U(1)'$ gauge symmetry. 
 In the secluded dark photon model~\cite{Huh:2007zw, Pospelov:2007mp, Chun:2010ve}, the new gauge boson $A'$ is kinetically mixed with the SM $U(1)$ gauge boson and that is how DM can interact with SM particles through the vector portal. 
 Whereas, for another model we explored, namely, $U(1)_{B-L}$, the new $U(1)'$ gauge symmetry stands for the conservation of the difference between the baryon number and the lepton number. 
 DM fermions are assumed to be charged under the $U(1)_{B-L}$ symmetry, while SM fermions are charged under $U(1)_{B-L}$ for obvious reason. 
 Similar logic holds for the $U(1)_{L_e - L_\mu} $ model where the new $U(1)'$ gauge symmetry stands for the conservation of the difference between the electronic lepton number and the muonic lepton number. 
 All of the models stated above are capable of depicting DM-SM interactions through the new $U(1)'$ gauge boson as the mediator.\footnote{For the impact of the interference between the standard and non-standard neutrino interactions in $U(1)$-extended models, see Ref.~\cite{Park:2023hsp} where it is shown that the interference can lead to a transition between non-standard interaction models in the energy range relevant to both dark matter and neutrino experiments.}
 We will work on the above mentioned benchmark models and show that the energy dependence of the cross section plays a crucial role in improving the constraints.
 
 We organize the subsequent discussion as follows. 
 In Section~\ref{sec:CR_composition_flux}, we briefly summarize the leading components of cosmic ray and estimate the corresponding fluxes. 
 We describe the benchmark models we choose to present in this work and discuss the existing constraints on the couplings for the models in Section~\ref{sec:Models}. 
 In Section~\ref{sec:BDM_flux}, we compute the flux of CRBDM from the knowledge of the fluxes of different cosmic-ray components. 
 We write down the expression for the calculation of event rate in Section~\ref{sec:event_rate}. 
 In Section~\ref{sec:constraint_eps}, we constrain the dark-sector coupling using the XENONnT data. 
 We translate the above mentioned bounds to the DM-electron interaction cross section as a function of DM mass for the corresponding models in Section~\ref{sec:cs_constraint}. 
 In Section~\ref{sec:attenuation}, we present a rough estimation of the upper bound on the DM-electron interaction cross section due to the shielding effect by the Earth crust. 
 Finally, we conclude in Section~\ref{sec:conclusion}. 
 For additional information, please refer to the Appendices, where we include the velocity distribution of CRBDM for some particular parameter sets (Appendix~\ref{sec:vel_dist}), the comparison of the contributions to the CRBDM fluxes by each individual components of cosmic rays (Appendix~\ref{sec:comp_BDM_flux}), and a simple analytical form for a quick estimation of cross section for the secluded dark-sector model (Appendix~\ref{sec:cs_est}).

\section{Cosmic Rays}
\label{sec:CR_composition_flux}

 As per our present understanding, cosmic rays  always consist of charged particles. 
 Additionally, we choose the definition of cosmic rays where neutrinos and gamma rays are excluded. 
 Based on the astrophysical origin, we can classify the cosmic rays into primary, secondary, tertiary rays and so on. 
 The particles, which are accelerated at astrophysical sources, are defined as primary cosmic rays. 
 Secondary ones consist of the particles produced in interaction between the interstellar gas and primary particles. 
 Further interaction between the interstellar gas and the secondary particles, produce tertiary rays and so on.
 The composition of cosmic rays near the Earth is inferred from the observations. 
 About $2 \%$ of the total observed cosmic rays are electrons and rest are baryonic particles. 
 Among baryonic component, about $87 \%$ are free protons, remaining cosmic rays consist of $12 \%$ helium nuclei and $1 \%$ heavier nuclei~\cite{Longair:1992ze}. 
 Please refer to Fig.~\ref{fig:CR_chart} for an overall picture at a glance. 
 While considering only the primary cosmic-ray particles, among nucleons $74 \%$ are free protons and $18 \%$ are in bound state inside helium nuclei~\cite{Workman:2022ynf}.
 
  For primary nucleons, the intensity is roughly parametrized as 
\bea  
I_N(E) \approx 1.8 \times 10^4 \left(\frac{E}{1~\rm{GeV}}\right)^{- \alpha}~\rm{\frac{nucleons}{m^2 \cdot s \cdot sr \cdot GeV}}\,,
\eea
where $E$ is the energy-per-nucleon and $\alpha = 2.7$.
To get the fluxes of the various components of cosmic ray, we follow Refs.~\cite{PhysRevD.98.030001,Boschini:2018zdv}. 
We use the parametrization of these fluxes described in Section~\ref{subsec:e_flux} and \ref{subsec:pHe_flux} in our further calculations.

\tikzset{
photon/.style={decorate, decoration={snake,amplitude=4pt, segment length=5pt}, draw=black},
particle/.style={draw=black, postaction={decorate}, decoration={markings,mark=at position .5 with {\arrow[draw=black]{>}}}},
antiparticle/.style={draw=black, postaction={decorate}, decoration={markings,mark=at position .5 with {\arrow[draw=black]{<}}}}
,gluon/.style={decorate, draw=black, decoration={coil,amplitude=4pt, segment length=5pt}},
process/.style={rectangle,
      draw=black,
      thick,
      text width=8em,
      align=center,
      rounded corners,
      minimum height=2em},
arrow/.style={thick,->,>=stealth}      
}
\begin{figure}
\begin{center}
\begin{tikzpicture}[thick,scale=1.0]
\node(CR) [process]{Cosmic Rays};
\node(Nuc) [process, left of=CR, xshift=-2cm, yshift=-2.0cm]{Protons + Nuclei \\ (98\%)};
\node(Elc) [process, right of=CR, xshift=2cm, yshift=-2.0cm]{Electrons \\ (2\%)};
\node(Proton) [process, right of=Nuc, xshift=-5cm, yshift=-2.0cm]{Protons \\ (87\%)};
\node(He) [process, right of=Nuc, xshift=-1cm, yshift=-2.0cm]{Helium Nuclei \\ (12\%)};
\node(Heavy) [process, right of=Nuc, xshift=3cm, yshift=-2.0cm]{Heavier Nuclei \\ (1\%)};
\draw [arrow] (CR) -- (Nuc);
\draw [arrow] (CR) -- (Elc);
\draw [arrow] (Nuc) -- (Proton);
\draw [arrow] (Nuc) -- (He);
\draw [arrow] (Nuc) -- (Heavy);
\end{tikzpicture}
\end{center}
\caption{Leading components of cosmic ray and corresponding composition~\cite{Longair:1992ze}.}
\protect\label{fig:CR_chart}
\end{figure}
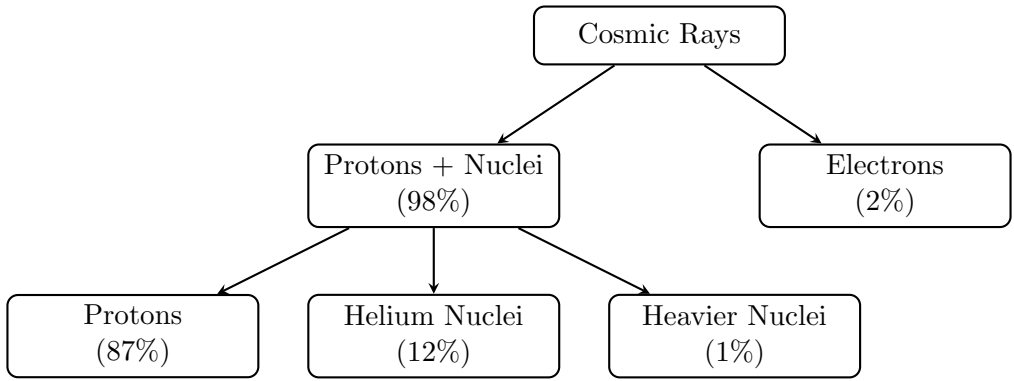

\subsection{Cosmic-Ray Electron Flux}
\label{subsec:e_flux}

 The cosmic-ray electron flux can be described by certain parametrization of the local interstellar spectrum~\cite{Boschini:2018zdv}, given as
\begin{align}
  \frac{d\Phi_e}{dT_e}(T_e) = \begin{cases}
     \mbox{\Large\( \frac{1.799 \times 10^{44}~ T_e^{-12.061}}{1+ 2.762 \times 10^{36}~ T_e^{-9.269} + 3.853 \times 10^{40}~ T_e^{-10.697}} \)} & \text{for $T_e < 6880$ MeV}\,, \\
   \mbox{\tiny\( \)} & \mbox{\tiny\( \)} \\
      3.259 \times 10^{10}~T_e^{-3.505} + 3.204 \times 10^{5}~T_e^{-2.620} & \text{for $T_e \geqslant 6880$ MeV}\,,
    \end{cases}
\label{Eq:CRe_flux_parameterization}
\end{align}
where $\frac{d\Phi_e}{dT_e}(T_e)$ is given in $\rm{\left(m^2 \cdot s \cdot sr \cdot MeV \right)^{-1}}$ and the kinetic energy of the cosmic-ray electrons ($T_e$) is in MeV. 
The above parametrization is in accordance with the data from Fermi-LAT~\cite{Fermi-LAT:2011baq,Fermi-LAT:2009yfs,Fermi-LAT:2010fit,Fermi-LAT:2017bpc}, AMS-02~\cite{AMS:2014gdf}, PAMELA~\cite{PAMELA:2011bbe,CALET:2017uxd}, and Voyager~\cite{Cummings:2016pdr,Stone:2013}.

\subsection{Cosmic-Ray Proton and Helium Fluxes} 
\label{subsec:pHe_flux}

 To estimate the cosmic-ray proton and helium fluxes, we follow Refs.~\cite{Boschini:2017fxq,DellaTorre:2016jjf} and use their analytical form of the functional dependence of the local interstellar proton and helium spectrum on rigidity (R). 
 The differential intensity in terms of rigidity is given by
 \begin{align}
  \frac{dI}{dR} \times R^{2.7} = \begin{cases}
     \sum_{i=0}^{5} a_i R^i & \text{for $R \leqslant 1$ GV}\,, \\
   \mbox{\tiny\( \)}  & \mbox{\tiny\( \)} \\
      \mbox{\Large\( b + \frac{c}{R} + \frac{d_1}{d_2 + R} + \frac{e_1}{e_2 + R} + \frac{f_1}{f_2 + R} + gR \)} & \text{for $R > 1$ GV}\,,
    \end{cases}
\label{Eq:CRp_flux_parameterization}
\end{align}
 where the parameter set are given in Table~\ref{Table:param}.
 The cosmic-ray proton and Helium fluxes are then obtained by the following relations
  \bea
  \frac{d\Phi_p}{dT_p}(T_p) = 4 \pi \frac{dR}{dT_p} \frac{dI}{dR}\,, \nonumber \\
  \frac{d\Phi_{He}}{dT_{He}}(T_{He}) = 4 \pi \frac{dR}{dT_{He}} \frac{dI}{dR}\,.
  \label{Eq:CRHe_flux_parameterization}
  \eea
These analytical forms of the fluxes stated above are useful in our analysis where DM is boosted by the electrons and/or nucleus depending on the model considered. 
We choose to work on the models described in Section~\ref{sec:benchmark}. 
Based on the expressions (Eqs.
(\ref{Eq:CRe_flux_parameterization}), (\ref{Eq:CRp_flux_parameterization}), and (\ref{Eq:CRHe_flux_parameterization})), we estimate the CRBDM flux in Section~\ref{sec:BDM_flux}.

 \begin{table}
 \begin{center}
 \begin{tabular}{|c |c c c c c c |}
 \hline \hline
 & {{$a_0$}} & {{$a_1$}} & {{$a_2$}} & {{$a_3$}} & {{$a_4$}} & {{$a_5$}} \\
 \hline 
  {{p}} & {{ $94.1$}} & {{$-831$}} & {{$0$}} & {{$16700$}} & {{$-10200$}} & {{$0$}} \\
  
  {{He}} & {{ $1.14$}} & {{$0$}} & {{$-118$}} & {{$578$}} & {{$0$}} & {{$-87$}} \\  
 \hline
 \end{tabular} 
 
  \begin{tabular}{|c | c c c c c c c c c|}
 \hline
 & {{$b$}} & {{$c$}} & {{$d_1$}} & {{$d_2$}} & {{$e_1$}} & {{$e_2$}} & {{$f_1$}} & {{$f_2$}} & {{$g$}} \\
 \hline 
  {{p}} & {{$10800$}} & {{$8590$}} & {{$-4230000$}} & {{$3190$}} & {{$274000$}} & {{$17.4$}} & {{$-39400$}} & {{$0.464$}} & {{$0$ }}\\
  
  {{He}} & {{$3120$}} & {{$-5530$}} & {{$3370$}} & {{$1.29$}} & {{$134000$}} & {{$88.5$}} & {{$-1170000$}} & {{$861$}} & {{$0.03$ }}\\  
 \hline \hline
 \end{tabular}
  \end{center}
 \caption{Parameters of the analytical fit of the local interstellar proton and Helium spectrum}
  \label{Table:param}
  \end{table}

\section{Models}
\label{sec:Models}

In this section, we describe the benchmark models to examine in this work and briefly discuss the existing constraints on their model parameters.
Our study considers models with an additional $U(1)'$, including the secluded dark photon, $U(1)_{B-L}$, and $U(1)_{L_e-L_\mu}$, to investigate the impact of cosmic-ray components on the CRBDM flux and the energy dependence of the cross section.

\subsection{Benchmark Models}  
\label{sec:benchmark}

 To provide an extensive analysis, we work on the following benchmark models. 
 The coupling to the SM leptons and/or baryons varies for different models and that yields variations in our final result as we will see in Section~\ref{sec:result}.
 
 \begin{itemize}
 
 \item Secluded dark sector (dark photon):
 We consider a hidden sector with a new $U(1)'$ gauge symmetry ($A'$ being the new gauge boson) and a Dirac fermion $\chi$,
 \bea
 \mathcal{L} &=& \mathcal{L}_{SM} + \bar{\chi} \left(i \slashed{\partial} - m_\chi \right) \chi - g_{\chi} \bar{\chi} \gamma_\mu \chi \hat{A'}^\mu \nonumber \\ &+& \frac{1}{2} m_{\hat{A'}}^2 \hat{A'}_\mu \hat{A'}^\mu  - \frac{1}{4} \hat{A'}_{\mu \nu} \hat{A'}^{\mu \nu} - \frac{\sin \varepsilon}{2} \hat{B}_{\mu \nu} \hat{A'}^{\mu \nu}
 ,.
 \eea 
 Here, the additional $U(1)'$ is assumed to be spontaneously broken which yields the mass $m_{\hat{A'}}$. 
 Due to the $U(1)_Y \times U(1)'$ kinetic mixing, the dark photon ($A'$) can couple to the SM fermions~\cite{Chun:2010ve,Fabbrichesi:2020wbt,Gu:2017gle}. 
 We follow Ref.~\cite{Chun:2010ve} to diagonalize away the kinetic mixing term and get the relevant interaction terms as follows
 \bea
 \mathcal{L} &\supset& {A'}_\mu \left[ g_{fL}^{A'}~\bar{f} \gamma^\mu P_L f + g_{fR}^{A'}~\bar{f} \gamma^\mu P_R f + g_{\chi}^{A'}~\bar{\chi} \gamma^\mu \chi \right] \nonumber \\
 &+& {Z}_\mu \left[ g_{fL}^{Z}~\bar{f} \gamma^\mu P_L f + g_{fR}^{Z}~\bar{f} \gamma^\mu P_R f + g_{\chi}^{Z}~\bar{\chi} \gamma^\mu \chi \right]\,,
 \eea 
 where $f$ being the SM fermions and the relevant couplings are taken from Ref.~\cite{Chun:2010ve}. 
 In this model, the charged leptons and baryons are coupled to the DM fermion ($\chi$) through the mediator ($A^{\prime}$) with the coupling strength proportional to the electromagnetic charge at the leading order. 
 In the case of DM-nuclear interaction, the cross section is therefore related to the DM-nucleon cross section by a factor which is proportional to the square of the number of protons inside the corresponding nucleus.
 
 \item $U(1)_{B-L}$: 
 The conservation of the difference between the baryon number and the lepton number is described by the new $U(1)'$ gauge symmetry. 
 DM fermions are considered to be charged under the new $U(1)'\equiv U(1)_{B-L}$ and SM fermions are already charged under the new gauge symmetry due to the baryonic and leptonic quantum numbers~\cite{Bauer:2018onh},
 \bea
\mathcal{L}_{B-L} \supset g_{B-L} \left[- \bar{l} \gamma^\mu A^{'}_{\mu} l + \frac{1}{3} \bar{q} \gamma^\mu A^{'}_{\mu} q \right] 
 - g_{\chi} \bar{\chi} \gamma_\mu \chi {A'}^\mu\,.
\eea
 In this model, all the leptons and the baryons are coupled to the DM fermions ($\chi$) through the mediator ($A^{\prime}$) via the lepton or baryon number. 
 Thus, in the case of DM-nuclear interaction, the cross section is enhanced from the DM-nucleon cross section by a factor which is proportional to the square of the total number of baryons inside the corresponding nucleus.

\item $U(1)_{L_{e} - L_{\mu}}$: 
 In the $U(1)_{L_e - L_\mu} $ model, the new $U(1)'$ gauge symmetry stands for the conservation of the difference between the electronic lepton number and the muonic lepton number~\cite{Bauer:2018onh},
\bea
\mathcal{L}_{L_{e} - L_{\mu}} \supset g_{L} \left[\bar{l}_e \gamma^\mu A^{'}_{\mu} l_e -\bar{l}_\mu \gamma^\mu A^{'}_{\mu} l_\mu  \right]- g_{\chi} \bar{\chi} \gamma_\mu \chi {A'}^\mu\,.
\eea
In the framework of this model, the DM cannot be boosted by CR protons or helium nuclei.
\end{itemize}

All of the models stated above are capable of depicting DM-SM interactions through the new $U(1)'$ gauge boson as the mediator. 
We use the above mentioned models to estimate the CRBDM flux (Section~\ref{sec:BDM_flux}) and the event rates (Section~\ref{sec:event_rate}) at XENONnT in the following discussions.

\subsection{Existing constraints on the model parameters}  
\label{sec:existing_constraints}

 In this subsection, we discuss the various existing constraints on the overall coupling between the DM and SM fermions for different values of the mass of the corresponding mediators. 
 Depending on the masses of the new $U(1)'$ gauge boson and the dark fermions, mainly two kinds of exclusion limits exist in the present literature~\cite{Inan:2021dir,Filippi:2020kii,Gninenko:2020hbd,Bauer:2018onh}.
 \begin{enumerate}
 \item For $m_\chi > m_{A'}/2 $, the dominant decay channel is $A' \to e^- e^+$. 
 The branching ratio of $A' \to e^- e^+$ is considered to be unity for obtaining the limits, and the process is termed as the visible decay mode of a dark photon.
 \item For $m_\chi < m_{A'}/2$, the dominant decay channel is $A' \to \chi \bar{\chi}$. 
 Now, the branching ratio of $A' \to \chi \bar{\chi}$ is considered to be unity for obtaining the limits, and the process is termed as the invisible decay mode of a dark photon.
 \item In literature, the semi-visible channels also exist but those channels are specifically dependent on the model parameters~\cite{Mohlabeng:2019vrz}.
 \end{enumerate}
 
 In our analysis, we initially constrain the coupling parameters between DM and SM particles for the benchmark models mentioned above and subsequently scattering cross sections. 
 To get an overview of the ruled-out region of the parameter space, we show the consolidated version of the strongest existing constraints obtained from different beam-dump and collider experiments (Fig.~\ref{fig:eps_constraint_invvis}) along with our results. 
 In the case of the invisible decay scenario, we have collected the exclusion limits from the Super Proton Synchrotron (SPS) experiments (NA64~\cite{Andreev:2021fzd,Banerjee:2019pds}, NA62~\cite{NA62:2019meo}), collider experiment (BaBar~\cite{Filippi:2019lfq,BaBar:2017tiz}), and electron beam-dump experiments (E787 and E949~\cite{Essig:2013vha}). 
 The strongest exclusion bounds for the visible decay of the dark photon have been gathered from SPS experiments (NA64~\cite{NA64:2019auh}, NA48~\cite{NA482:2015wmo}, WASA~\cite{WASA-at-COSY:2013zom}), collider experiments (BESIII~\cite{BESIII:2017fwv}, Babar~\cite{Filippi:2019lfq,BaBar:2014zli}, LSND~\cite{LSND:1997vqj}, HADES~\cite{HADES:2013nab}, KLOE~\cite{KLOE-2:2012lii,KLOE-2:2014qxg,Anastasi:2015qla}), electron beam-dump experiments (E141~\cite{Fabbrichesi:2020wbt,Riordan:1987aw}, E137~\cite{Fabbrichesi:2020wbt,Bjorken:1988as,Batell:2014mga,Marsicano:2018krp}, E774~\cite{Bross:1989mp}), proton beam-dump experiments(nu-Cal~\cite{Blumlein:2011mv,Blumlein:2013cua}, CHARM~\cite{Gninenko:2012eq}), and electron-nucleus fixed target scattering experiments (APEX~\cite{APEX:2011dww}, A1~\cite{Merkel:2014avp}). 
For more details, please refer to Refs.~\cite{Inan:2021dir,Filippi:2020kii,Gninenko:2020hbd,Bauer:2018onh} and Section~\ref{sec:constraint_eps}.

\section{Boosted Dark Matter Flux}
\label{sec:BDM_flux}

 In this section, we estimate the BDM flux while taking into account the potential candidates of cosmic rays as boosting agent of cold DM particles, namely, cosmic-ray electrons, protons, and Helium nuclei. 
 We calculate the CRBDM flux using the knowledge of cosmic-ray electron, proton, and Helium fluxes summarized in Section~\ref{sec:CR_composition_flux}.
 
For the scattering between cosmic ray and DM, the energy transfer to the cold DM from the CR electrons/protons/Helium nuclei is given by~\cite{Ema:2018bih,Bringmann:2018cvk,Cappiello:2019qsw,Bardhan:2022bdg,Ghosh:2021vkt}
    \bea
  T_{\chi} &=&  T_{\chi}^{\rm max}  \left( \frac{ 1 - \cos \theta }{2}\right) \nonumber \\ 
  {\rm with} \quad T_{\chi}^{\rm max} &=& \frac{\left(T_i\right)^2+ 2 T_i m_i  }{T_i+ \left(m_i +m_{\chi} \right)^2/\left(2 m_{\chi} \right)}\,, 
  \label{Eq:T-chi-max} 
  \eea
 where $T_i$ is the kinetic energy of the cosmic-ray particle $i$ and $\theta$ is the scattering angle at the center of momentum frame.
 Solving Eq.~(\ref{Eq:T-chi-max}), we get the minimum required energy of the cosmic-ray particles to produce BDM with a certain amount of kinetic energy $T_\chi$:
 \bea
  T_i^{\rm min} = \left( \frac{T_{\chi}}{2} -m_i \right) \left[ 1 \pm \sqrt{ 1 +  \frac{2 T_{\chi}}{ m_{\chi}} \frac{\left(m_i +m_{\chi}\right)^2}{\left( 2m_i - T_{\chi} \right)^2}} \right]\,, 
  \label{Eq:T-nu-min}
  \eea
 where the $+$ and $-$ signs in Eq.~(\ref{Eq:T-nu-min}) are applicable for $T_\chi > 2 m_i$ and $T_\chi < 2 m_i$, respectively.

 Finally, the CRBDM flux takes the form   
 \bea
 \frac{d\Phi_\chi}{dT_\chi} &=& D_{eff} \times \frac{\rho_\chi^{\rm{local}}}{m_\chi}~  \Biggl[ \int^{\infty}_{T_e^{\rm min}} dT_e ~ \frac{d\Phi_e}{dT_e}~ \frac{d \sigma_{\chi e}}{dT_\chi} \nonumber \\ &+& \int^{\infty}_{T_p^{\rm min}} dT_p ~ \frac{d\Phi_p}{dT_p}~ \frac{d \sigma_{\chi p}}{dT_\chi} G_{p}^2(2m_\chi T_\chi) \nonumber \\ &+& \int^{\infty}_{T_{\rm He}^{\rm min}} dT_{\rm He} ~ \frac{d\Phi_{\rm He}}{dT_{\rm He}}~ \frac{d \sigma_{\chi {\rm He}}}{dT_\chi} G_{\rm He}^2(2m_\chi T_\chi) \Biggr] \nonumber \\
 \label{Eq:DMflux-wrt-T-full}
 \eea
 with 
 \bea
 \frac{d \sigma_{\chi i}}{dT_\chi} = {g_{i}^{A'}}^2 {g_{\chi}^{A'}}^2 \frac{2m_\chi \left(m_i + T_i \right)^2 - T_\chi \left\lbrace \left( m_i + m_\chi \right)^2 + 2 m_\chi T_i  \right\rbrace + m_\chi T_\chi^2}{4 \pi \left(2 m_i T_i + T_i^2 \right) \left(2 m_\chi T_\chi + m_{A'}^2 \right)^2}\quad (i=e,p,{\rm He})\,, \nonumber \\
 \label{eq:diff_cs_p}
 \eea
 where $D_{eff}$ is an effective distance out to which we take into account CRs as the source of a possible high-velocity tail in the DM velocity distribution, $\rho_\chi^{\rm{local}}$ is the local energy density of DM, $G_i$ be the nucleon electromagnetic form factors. 
 We include this form factor for the hadronic elastic scattering following Ref.~\cite{Bringmann:2018cvk,Perdrisat:2006hj,Lei:2020mii},
 \bea
 G_i\left(q^2\right) = \left(1+ \frac{q^2}{\Lambda_i^2} \right)^{-2}\,,
 \eea
 where $\Lambda_p= 770~\rm{MeV}$, $\Lambda_{He}= 410~\rm{MeV}$, and $q$ is the momentum transfer. 
 We evaluate the BDM fluxes after including this form factor, and do the same calculation after including the Helm form factor~\cite{Cho:2020mnc}. 
 The obtained fluxes are comparable. 
 In the point-like limit, the differential scattering cross section can be related by 
 \bea
 \frac{d\sigma_{\chi i}}{d\Omega} = \frac{d\sigma_{\chi i}}{d\Omega}\Bigg|_{q^2=0} G_i\left(2m_\chi T_\chi\right)\,.
 \eea

 Overall couplings will be a bit different for all three models, and thus for $L_{e} - L_{\mu}$ model there will be no contribution from the baryonic sector, i.e., for boosting the DM, proton and helium nuclei will not take part. 
 For different considered models, couplings in Eq.~(\ref{eq:diff_cs_p}) are as follows.
 \begin{itemize}
 \item Secluded dark sector: 
$\abs{g_{e}^{A'}} = \abs{g_{p}^{A'}} =  e \varepsilon \cos \theta_W$, $\abs{g_{He}^{A'}} = 2 \abs{g_{p}^{A'}}$, $g_{\chi}^{A'} = g_{\chi}$ with $\theta_W $ being the Weinberg angle.
 \item $U(1)_{B-L}$: $\abs{g_{e}^{A'}} = \abs{g_{p}^{A'}} =  g_{B-L}$, $\abs{g_{He}^{A'}} = 4 \abs{g_{p}^{A'}}$, $g_{\chi}^{A'} = g_{\chi}$.
 \item $U(1)_{L_e - L_\mu}$: $\abs{g_{e}^{A'}} =  g_{L}$, $g_{\chi}^{A'} = g_{\chi}$.
 \end{itemize}

\begin{figure*}
 \centering
 \includegraphics[width=0.49 \textwidth]{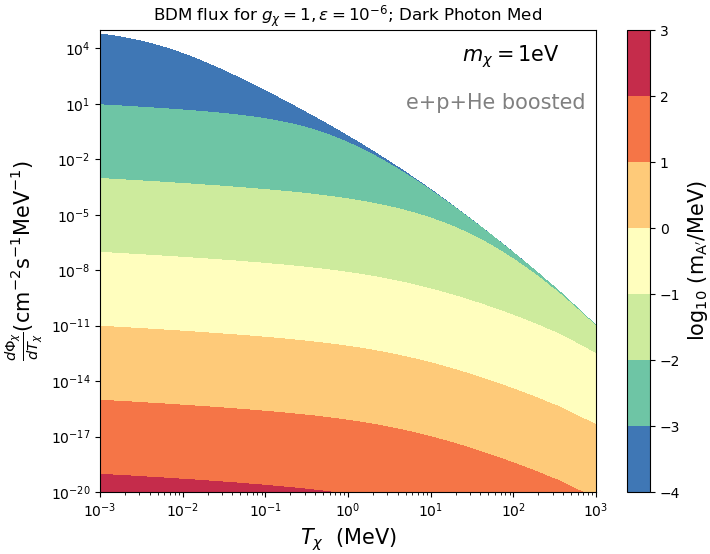}
 \includegraphics[width=0.49 \textwidth]{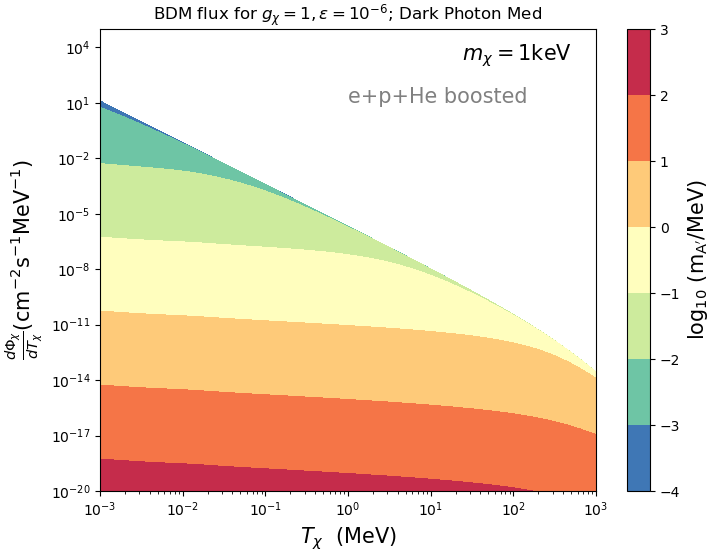}\\
 \includegraphics[width=0.49 \textwidth]{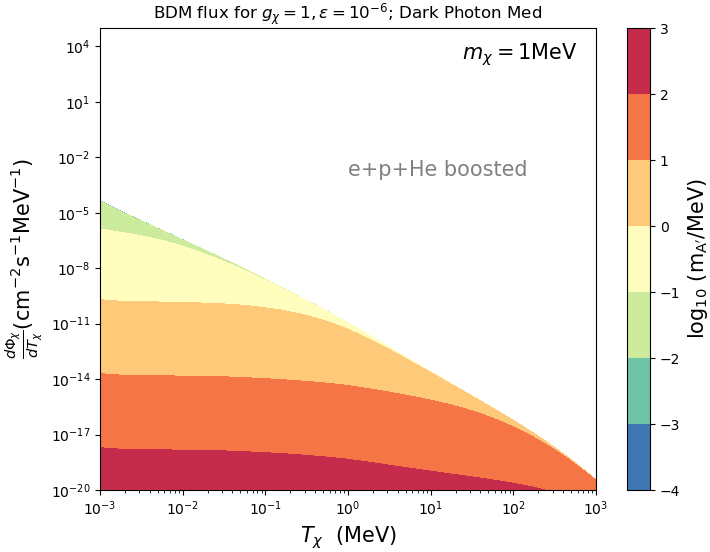}
 \includegraphics[width=0.49 \textwidth]{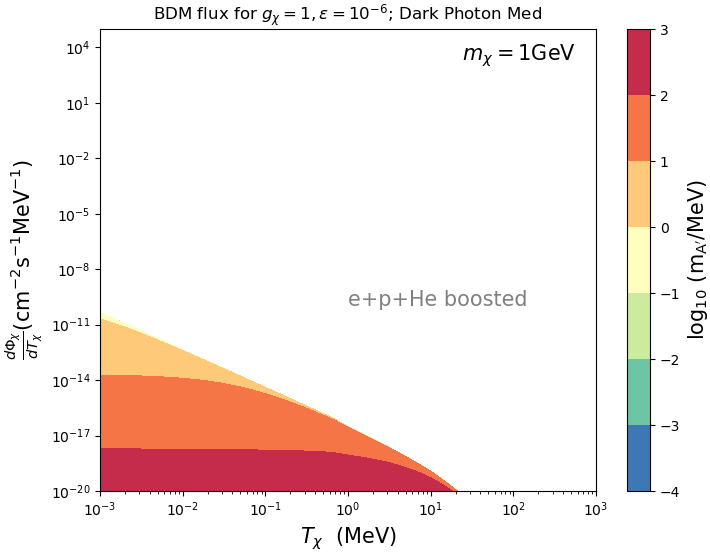}
 \caption{Cosmic-ray electron, proton and Helium nuclei boosted DM flux for the secluded dark photon model with $g_\chi = 1, \varepsilon = 10^{-6}$ and $m_\chi = 1~\rm{eV}, 1~\rm{keV}, 1~\rm{MeV}, 1~\rm{GeV}$, respectively.
 }
 \label{fig:flux_eps}
 \end{figure*}

We start with the secluded dark photon model to calculate cosmic-ray electron-boosted DM flux which is in agreement with that in Ref.~\cite{Cao:2020bwd}. 
Then we calculate the boosted DM flux and show it in Fig.~\ref{fig:flux_eps} taking all the contributions from cosmic-ray electrons, protons and helium nuclei. 
In this model, the boost contribution due to the helium nuclei is proportional to the number of protons inside the nucleus.
Similar flux estimation has been done for the $U(1)_{B-L}$ model and the $U(1)_{L_e - L_\mu} $ model. 
For a detailed discussion on the latter models, please refer to Appendix~\ref{sec:comp_BDM_flux}. 
For the $U(1)_{L_e - L_\mu} $ model, DM gets the boost only from the cosmic-ray electron, whereas for the $U(1)_{B-L}$ model, the boost contributions come from electrons, protons, and helium nuclei. 
Unlike the secluded dark sector, in the $U(1)_{B-L}$ model, the boost contribution due to the helium nuclei is proportional to the total baryon number of the nucleus. 
For the secluded dark photon model  with a reference parameter set of $g_\chi = 1, \varepsilon = 10^{-6}$, we plot the flux of the CRBDM in Fig.~\ref{fig:flux_eps}.
 
There are various interesting features in the CRBDM flux distribution. 
The interface of the two differently colored regions indicates the CRBDM flux for a particular value of $m_{A^{\prime}}$ which is denoted by the colorbar. 
From Fig.~\ref{fig:flux_eps}, we can find that for each $m_\chi$ there is a threshold of $m_{A^{\prime}}$ such that a fixed $m_\chi$ produces identical CRBDM flux distribution irrespective of the value of $m_{A^{\prime}}$ below the threshold. 
This phenomenon has been reflected in each panel as gradual squeezing of the particular colored region into a line, as one moves across the panels, i.e., from lower $m_\chi$ to higher $m_\chi$. 
For example, the panels of Fig.~\ref{fig:flux_eps} shows the gradual squeezing behaviour of the blue colored region into a line if one moves from $m_\chi$ = 1 eV to $m_\chi$ = 1 GeV.
In addition, we note that as $m_\chi$ increases the threshold value also increases, e.g., the BDM flux distribution corresponding to $m_{A'} = 10^{-4}$ and $10^{-3}$ MeV is significantly different for $m_\chi = 1~\rm{eV}$ but almost similar for $m_\chi = 1~\rm{keV}$. 
Analogously, for $m_\chi = 1~\rm{MeV}$ and $1~\rm{GeV}$ the BDM flux distribution becomes independent of $m_{A'}$ when considering $m_{A'} < 10^{-2}$ MeV and $1$ MeV, respectively. 
For very high $m_{A'}$, the BDM flux becomes nearly independent of $m_\chi$. 
In the subsequent section, we will see that these features induce interesting behaviors in exclusion limits as well.

\section{Results}  
\label{sec:result}

In this section, we calculate the expected event rate of CRBDM in direct detection experiments.
Utilizing the expected event rate and the XENONnT data, we estimate the constraints on the new couplings of the benchmark models which are later translated to the bounds on the DM-electron scattering cross section.
In addition, we provide an estimation procedure of the upper bound on the DM-electron interaction cross section due to the Earth shielding effect.

\subsection{Differential event rate} 
\label{sec:event_rate}

 With the knowledge of the CRBDM flux discussed in the previous section, we can estimate the differential event rate for the electron recoil spectrum in XENONnT~\cite{Das:2021lcr, Bardhan:2022bdg, Ghosh:2021vkt},
 \bea
 \frac{dR}{dE_R} = \frac{Z_{\rm Xe}}{m_{\rm Xe}} \int_{T_\chi^{\rm min}}^\infty dT_\chi \frac{d\Phi_\chi}{dT_\chi} \frac{d\sigma_{\chi e}}{dE_R}\,,
 \label{Eq:diff_rate}
 \eea
 where $Z_{\rm Xe}$ is the effective atomic number of xenon and $m_{\rm Xe}$ is the mass of a single xenon atom. 
 $Z_{\rm Xe}$ has been taken as $40$ to incorporate the fact that all the xenon atoms are not available for ionization~\cite{Das:2021lcr}.

  We do chi-square analysis to obtain the exclusion limits in Section~\ref{sec:constraint_eps} and the subsequent discussions. 
  In addition, we use a convolution along with the electron recoil spectrum given in Eq.~(\ref{Eq:diff_rate}) as suggested by Ref.~\cite{XENON:2020rca}, which is basically a Gaussian detector response function with a width given as
  \bea
  \sigma (E) = a \sqrt{E} + b E\,,
  \eea
  where $a = 0.31~\rm{\sqrt{keV}}$ and $b = 0.0037$.

\subsection{Constraints on the coupling}  
\label{sec:constraint_eps}
 
\begin{figure*}
 \centering
 \includegraphics[width=0.49 \textwidth]{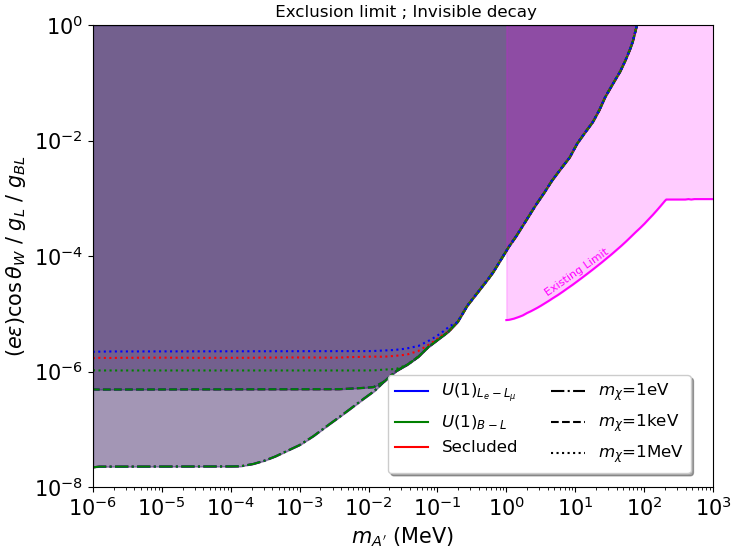}
 \includegraphics[width=0.49 \textwidth]{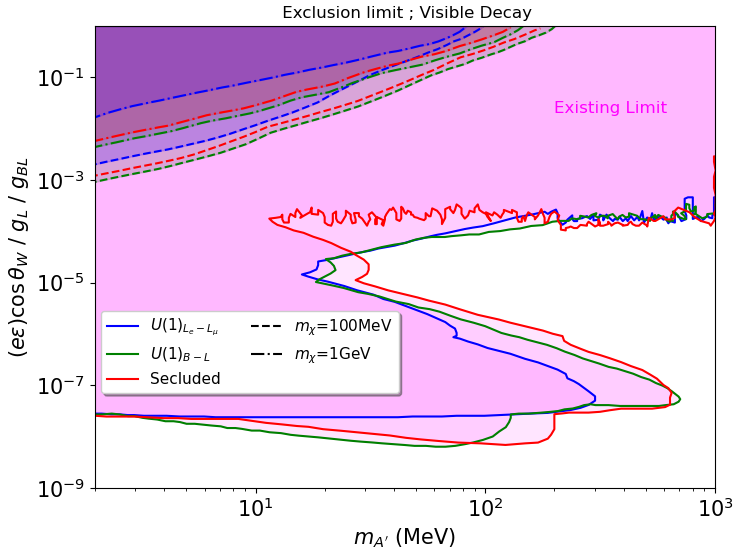}
 \caption{$95\%$ CL exclusion limits from CRBDM in the coupling vs $m_{A^{\prime}}$ plane: (left) the invisible decay scenario and (right) the visible decay scenario.
 We compare the three benchmark models for five values of $m_\chi$ with $g_\chi = 1$. 
 Shaded regions surrounded by solid lines show existing limits.}
 \label{fig:eps_constraint_invvis}
 \end{figure*}

We constrain the overall coupling for all the three models mentioned previously, using the XENONnT electron-recoil data~\cite{XENON:2022ltv}.\footnote{Actually, the limit obtained using XENON1T data~\cite{XENON:2020rca} is not significantly different from the XENONnT exclusion limit.} 
In Fig.~\ref{fig:eps_constraint_invvis}, we show the exclusion limit in the coupling vs $m_{A^{\prime}}$ plane for the scenarios of the invisible and the visible decay of dark photon by setting $g_\chi = 1$. 
We also show the existing limits for comparison~\cite{Bauer:2018onh}. 
In Ref.~\cite{Bauer:2018onh}, the visible decay bounds have small variations for three different models as shown in the right panel of Fig.~\ref{fig:eps_constraint_invvis}. 
For the invisible decay case, we show the existing exclusion limit in magenta. 
The bounds on the invisible decay of the hidden photon are from NA64~\cite{Andreev:2021fzd,Banerjee:2019pds}, NA62~\cite{NA62:2019meo}, Babar~\cite{Filippi:2019lfq,BaBar:2017tiz}, E787, and E949~\cite{Essig:2013vha}. 
The exclusion limits on the decay of dark photons into the pair of electron-positron ($e^+ e^-$) are from Babar~\cite{Filippi:2019lfq,BaBar:2014zli}, NA64~\cite{NA64:2019auh}, NA48~\cite{NA482:2015wmo}, E141~\cite{Fabbrichesi:2020wbt,Riordan:1987aw}, E137~\cite{Fabbrichesi:2020wbt,Bjorken:1988as,Batell:2014mga,Marsicano:2018krp}, E774~\cite{Bross:1989mp}, nu-Cal~\cite{Blumlein:2011mv,Blumlein:2013cua}, CHARM~\cite{Gninenko:2012eq}, KLOE~\cite{KLOE-2:2012lii,KLOE-2:2014qxg,Anastasi:2015qla}, WASA~\cite{WASA-at-COSY:2013zom}, HADES~\cite{HADES:2013nab}, APEX~\cite{APEX:2011dww}, A1~\cite{Merkel:2014avp},BESII~\cite{BESIII:2017fwv}, and LSND~\cite{LSND:1997vqj}. 
Our invisible limits in Fig.~\ref{fig:eps_constraint_invvis} correspond to the scenario when there is no provision for the new $U(1)'$ gauge boson ($A'$) to decay into the pair of electron-positron. 
Depending on the mass of DM ($m_\chi$) and the mass of the new $U(1)'$ gauge boson ($m_{A'}$), $A'$ can or cannot decay into a pair of DM particles ($\chi \bar{\chi}$). 
The behavior of the exclusion limit plot is consistent with variation of the BDM flux distribution.

As discussed earlier in Section~\ref{sec:BDM_flux}, there is a threshold of $m_{A'}$ for a given $m_\chi$, below which the CRBDM flux distribution almost is independent of $m_{A'}$, 
This behavior is reflected to the limits in Fig.~\ref{fig:eps_constraint_invvis}. 
The exclusion limit becomes constant below the threshold of $m_{A'}$ for a fixed $m_\chi$, and the threshold also depends on $m_\chi$ as the flux does.
On the other hand, above the threshold the BDM flux decreases with increasing $m_{A'}$, and therefore the exclusion limit becomes weaker. 
For higher $m_{A'}$, the exclusion limit becomes nearly independent of $m_\chi$.
Note that only the fluxes for larger than $T_{\chi}^{\rm min}$ (from Eq.~(\ref{Eq:diff_rate})) contribute to the event rate and lower $m_\chi$ DM is mostly boosted by CR electron as can be seen from Appendix~\ref{sec:comp_BDM_flux}. 

\begin{figure*}
 \centering
 \includegraphics[width=0.49 \textwidth]{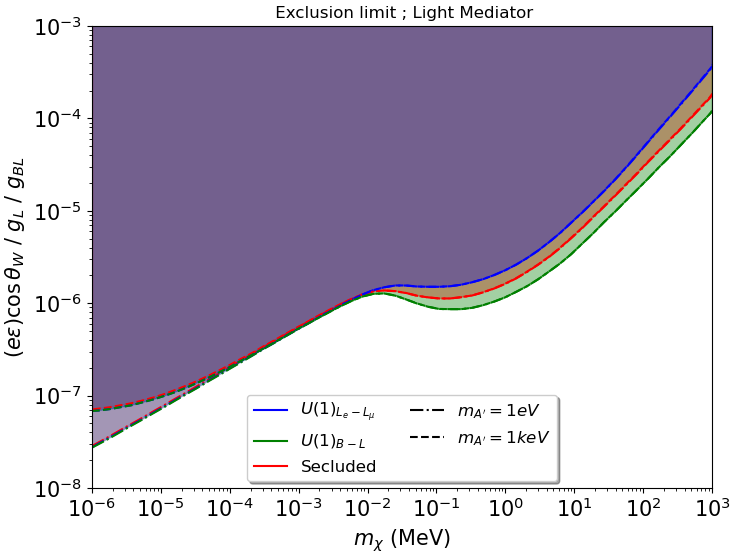}
 \includegraphics[width=0.49 \textwidth]{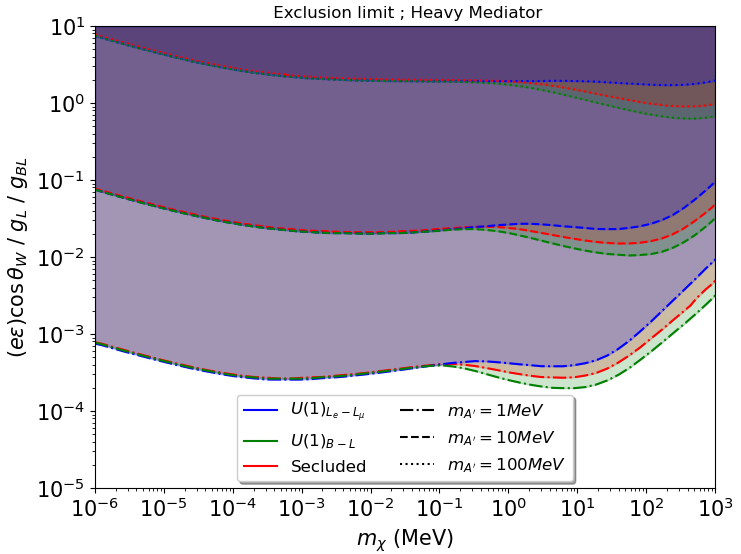} \caption{$95\%$ CL exclusion limits in the coupling vs $m_\chi$ plane for the three benchmark models and for five values of $m_\chi$ with $g_\chi = 1$: (left) the light mediator scenario and (right) the heavy mediator scenario.
 }
 \label{fig:coup_constraint_LMHM}
 \end{figure*}

 In Fig.~\ref{fig:coup_constraint_LMHM}, we show the  exclusion limit for $g_\chi=1$ in the coupling vs $m_{\chi}$ plane for both of the light and heavy mediator scenarios. 
 We show the exclusion region for all the three models, namely, secluded dark photon, $U(1)_{B-L}$,  and $U(1)_{L_{e} - L_{\mu}}$. 
 Similar explanation can be drawn for Fig.~\ref{fig:coup_constraint_LMHM} using the flux results as done for Fig.~\ref{fig:eps_constraint_invvis}. 
 In the left panel, there is a small separation in the exclusion limits at low $m_\chi$ for $m_{A'}$=1 eV and 1 keV but at higher $m_\chi$ these two limits got merged. 
 This is because for higher $m_\chi$, the BDM flux distributions corresponding to $m_{A'}$=1 eV and 1 keV are identical but at lower $m_\chi$ there is a little difference. 
 In the right panel, we can see the exclusion limits tend to become nearly independent of $m_\chi$ as $m_{A'}$ increases which is consistent with results shown in Fig.~\ref{fig:eps_constraint_invvis}. 
 Depending on the mass of the mediator the above mentioned nearly constant lines shifts up or down. 
 For low $m_\chi$, the exclusion limits are same for all of the three benchmark models as the leading contribution comes from CR electron flux; but for higher $m_\chi$, the BDM flux contributions due to proton and helium nuclei become comparable and thus the splittings among limits on different models appear noticeably. 
 The starting point of the splitting depends on $m_{A'}$.
 
\subsection{Constraints on the scattering cross section} 
\label{sec:cs_constraint}

Instead of working with energy independent constant cross section, we choose to work on the energy dependent one taking the three benchmark models discussed in Section~\ref{sec:Models}. 
Given a specific model, the differential cross section depends on the DM mass and the kinetic energies of both DM and the electron.
Moreover, it depends on the momentum transfer in the collision,  and the DM-electron scattering cross section is conventionally normalized to $\bar{\sigma}_{\chi e}$ with the following definitions~\cite{Cao:2020bwd,Essig:2011nj}:
\bea
\bar{\sigma}_{\chi e} &=& \frac{\mu_{\chi e}^2~\overline{\abs{\mathcal{M}_{free}(\alpha m_e)}^2}}{16 \pi m_\chi^2 m_e^2}\,, \\ 
\overline{\abs{\mathcal{M}_{free}}^2} &=& \overline{\abs{\mathcal{M}_{free}(\alpha m_e)}^2} \times \abs{F_{DM}(q)}^2\,,
\eea
where the form factor is given by (at the detector)
\bea
\abs{F_{DM}(q=\sqrt{2 m_e E_R})}^2 &=& \frac{(\alpha^2 m_e^2 + m_{A'}^2)^2}{(2 m_e E_R + m_{A'}^2)^2} \nonumber \\
&\times& \frac{2 m_e (m_\chi + T_\chi)^2 - E_R \left[ (m_\chi + m_e)^2+ 2 m_e T_\chi\right] + m_e E_R^2}{2 m_e m_\chi^2} \,.
\eea 
%
In the non-relativistic limit $E_R, T_\chi \ll m_e$,
\bea
\abs{F_{DM}(q)}^2 = \frac{(\alpha^2 m_e^2 + m_{A'}^2)^2}{(q^2 + m_{A'}^2)^2} \,.
\eea 
which is simplified to $\abs{F_{DM}(q)} = 1$ in the heavy-mediator limit and $\abs{F_{DM}(q)} \sim \frac{1}{q^2}$ in the light-mediator limit.
With these definitions, the normalized DM-electron scattering cross section is estimated as
%
\bea
\overline{\abs{\mathcal{M}_{free}(\alpha m_e)}^2} &=& \frac{16 {g_{e}^{A'}}^2  {g_{\chi}^{A'}}^2 m_e^2 m_\chi^2}{(\alpha^2 m_e^2 + m_{A'}^2)^2}\,, \nonumber \\
\bar{\sigma}_{\chi e} &=& \frac{{g_{e}^{A'}}^2{g_{\chi}^{A'}}^2 \mu_{\chi e}^2}{\pi (\alpha^2 m_e^2 + m_{A'}^2)^2}\,, \nonumber \\
\implies \bar{\sigma}_{\chi e} &=& \begin{cases}
    \frac{{g_{e}^{A'}}^2{g_{\chi}^{A'}}^2 \mu_{\chi e}^2}{\pi (\alpha^2 m_e^2)^2} ~~~\text{for a light mediator,}   \\
        \frac{{g_{e}^{A'}}^2{g_{\chi}^{A'}}^2 \mu_{\chi e}^2}{\pi (m_{A'}^2)^2} ~~~\text{for a heavy mediator}.
        \end{cases}
\label{eq:sigma_from_coupling}        
\eea

\begin{figure*}
 \centering
 \includegraphics[width=0.49 \textwidth]{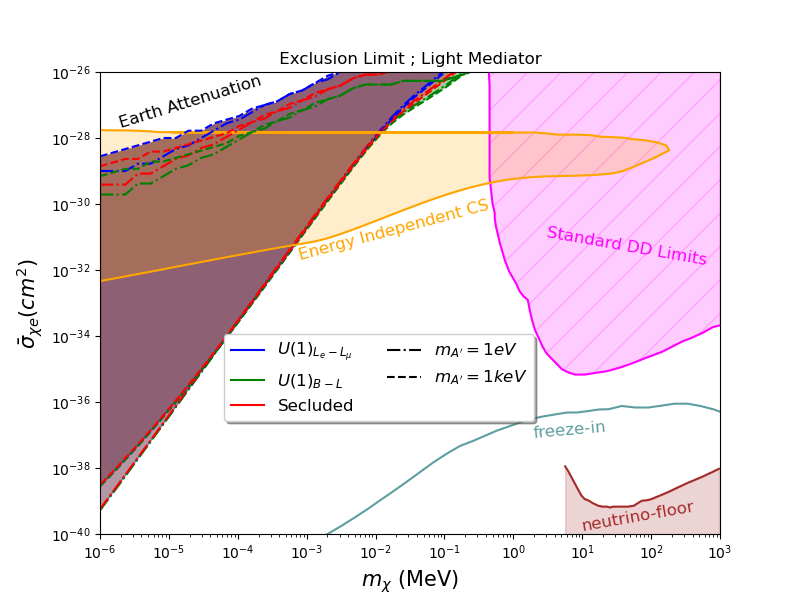}
 \includegraphics[width=0.49 \textwidth]{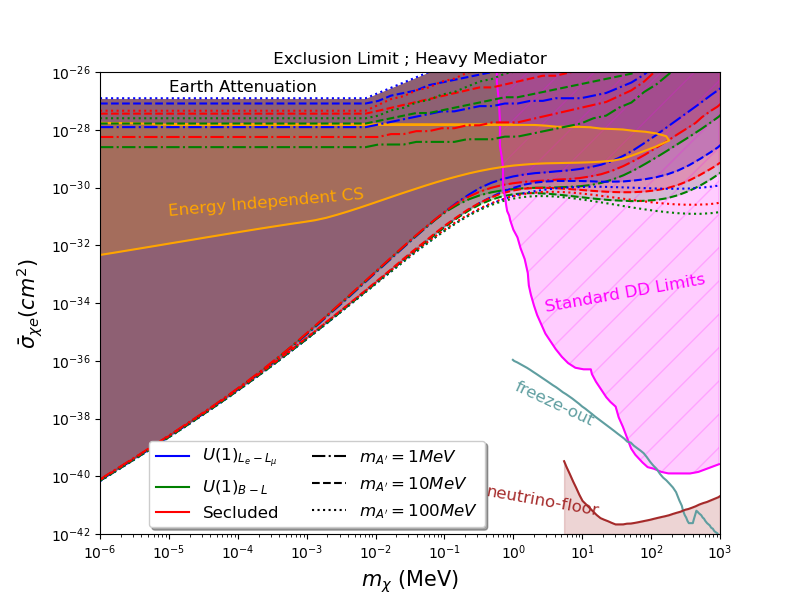}
 \caption{Constraints of DM-electron scattering cross section for $g_\chi = 1$ as a function of $m_\chi$ (Left) for the light mediator scenario and (right) the heavy mediator scenario. 
 The limit~\cite{Bardhan:2022bdg} for constant cross section scenario and the existing direct detection limits are included for comparison.
 The parameters which provide the observed DM relic abundance are also shown.
}
 \label{fig:sigmae_constraint_LMHM}
 \end{figure*}

We are now in the position to find the exclusion limit for the DM-electron scattering cross section from the limit for the couplings obtained in Section~\ref{sec:constraint_eps}.
 Fig.~\ref{fig:sigmae_constraint_LMHM} shows our exclusion limit for the $\bar{\sigma}_{\chi e}$ as a function $m_\chi$ for the light and the heavy mediator scenarios. 
 For comparison, we include the existing direct detection limits as a magenta shaded region, which are taken from SENSEI~\cite{SENSEI:2020dpa}, EDELWEISS~\cite{EDELWEISS:2020fxc}, PandaX-II~\cite{PandaX-II:2021nsg}, XENON100~\cite{Essig:2017kqs}, XENON1T~\cite{XENON:2019gfn}, SuperCDMS~\cite{SuperCDMS:2020ymb}, DAMIC~\cite{DAMIC:2019dcn}, and DarkSide-50~\cite{DarkSide:2018ppu}.
 The parameters that yield the DM relic abundance consistent with the observation are also shown: freeze-in~\cite{Griffin:2019mvc,Taufertshofer:2023rgq,Bellomo:2022qbx} for the light mediator scenario and freeze-out~\cite{Essig:2011nj} for the heavy mediator scenario. 
 To estimate the upper bound in order to take care of the Earth shielding effect, we compute the attenuated flux and find out the value of cross section for which it is significantly low. 
 More details of the Earth shielding effect is discussed in the following section.

Behavior of the exclusion limits on the cross section follows the tendency of the exclusion limits on the coupling. 
In both the left and the right panels of Fig.~\ref{fig:sigmae_constraint_LMHM}, we can see that, for the low $m_\chi$ the lower limits for all the three models get merged, but at higher $m_\chi$ they get split. 
This pattern is consistent with Fig.~\ref{fig:coup_constraint_LMHM} and follows from the behavior of BDM flux as discussed in Section~\ref{sec:constraint_eps}. 
However, the upper limits for all the three models never get merged. 
This is due to the different Earth shielding effect for the different models caused by the particles present inside the Earth crust; while travelling through the Earth crust towards the detector, BDM interacts only with electrons for the $U(1)_{L_e - L_\mu} $ model, only with the charged particles for the secluded dark photon model, and with all the baryons and leptons for the $U(1)_{B-L}$ model. 
In addition, we note that the splitting pattern of the lower limit for $m_{A^{\prime}}=1$ eV and $m_{A^{\prime}}=1$ keV is very similar for Fig.~\ref{fig:coup_constraint_LMHM} and Fig.~\ref{fig:sigmae_constraint_LMHM}, because for the light mediator the dependence of $\bar{\sigma}_{\chi e}$ on $m_{A^{\prime}}$ is negligible as seen from Eq.~(\ref{eq:sigma_from_coupling}). 
On the other hand, for the heavy mediator case, the dependence of $\bar{\sigma}_{\chi e}$ on $m_{A^{\prime}}$ becomes stronger, which causes the lower limits to get merged at low $m_\chi$ and only to get split at higher $m_\chi$. 
This behavior is also consistent with the right panel of Fig.~\ref{fig:coup_constraint_LMHM} where it can be seen that the bounds for different $m_{A^{\prime}}$ are almost parallel for low $m_\chi$ and the slopes change only at higher $m_\chi$. 
The slopes depend on the value of $m_{A^{\prime}}$; higher the $m_{A^{\prime}}$, flatter the bound is.

\subsection{Attenuation of BDM flux due to the Earth shielding effect} 
\label{sec:attenuation}

 The exact analysis for the upper bound due to the attenuation effect requires monte carlo simulation which is beyond the scope of this work. 
 For an alternative approach, which can give quite good approximate estimation, we basically follow Ref.~\cite{Emken:2018run}.
  The kinetic energies of a BDM particle before and after travelling through the Earth to the detector at depth $d$, are related via the following expression 
 \bea
 T_\chi^{\rm final}(d) = T_\chi^{\rm initial} \exp\Biggl[- \sum_N n_N \frac{2 \mu_{\chi N}^2 \sigma_{\chi N}}{m_N m_\chi} d - n_e \frac{2 \mu_{\chi e}^2 \sigma_{\chi e}}{m_e m_\chi} d \Biggr]\,,
 \label{eq:attenuated_KE}
 \eea
 where $n_N$ and $n_e$ are the number densities of nucleus species $N$ and electron in the Earth crust, respectively. 
 For the number densities of the nuclear species, we follow Ref.~\cite{Emken:2017qmp}~\footnote{$n_N = 2.7~\rm{g/cm^3} \cdot f_N \cdot N_A /A_N$ where $f_N, A_N$ are the mass fraction and the mass number of the species $N$ and $N_A$ is Avogadro's number.}. 
 On the other hand, we assume a constant number density of the electron, $n_e = 8 \times 10^{23}~\rm{cm^{-3}}$~\cite{Ema:2018bih}. 
 The DM-nucleus scattering cross section is related to the DM-nucleon scattering cross section via 
  \begin{align}
  \sigma_{\chi N} = \begin{cases}
     \sigma_{\chi n} A_N^2 \frac{\mu_{\chi N}^2}{\mu_{\chi n}^2} & \text{for the $U(1)_{B-L}$ model}\,, \\
   \mbox{\tiny\( \)}  & \mbox{\tiny\( \)} \\
      \sigma_{\chi n} Z_N^2 \frac{\mu_{\chi N}^2}{\mu_{\chi n}^2} & \text{for the secluded dark photon model}\,.
    \end{cases}
\label{Eq:nuclei_from_nucleon}
\end{align}
 Then we can estimate the attenuated CRBDM flux by the following expression~\cite{Bringmann:2018cvk}
 \bea
 \frac{d\Phi_\chi}{dT_\chi^{\rm final}}(d) = \frac{d\Phi_\chi}{dT_\chi^{\rm initial}}(d=0) \times \frac{dT_\chi^{\rm initial}}{dT_\chi^{\rm final}}\,.
 \eea
 Here $\frac{d\Phi_\chi}{dT_\chi^{\rm initial}}$ needs to be evaluated at $T_\chi^{\rm initial}$ where $T_\chi^{\rm initial}$ along with $\frac{dT_\chi^{\rm initial}}{dT_\chi^{\rm final}}$ can be evaluated using Eq.~(\ref{eq:attenuated_KE}) for any known $T_\chi^{\rm final}$.

 We put the modified flux, after considering the attenuation due to Earth scattering, into the event rate equation, Eq.~(\ref{Eq:diff_rate}), to evaluate the final differential event rate for electronic recoil. 
  We compute the differential event rate at the detector threshold energy of recoiled electron $\frac{dR}{dE_R}(E_R = E_R^{\rm th})$ for different values of constant cross section ($\bar{\sigma}_{\chi e} $). 
  Then the attenuation bound (upper limit on $\bar{\sigma}_{\chi e} $) is the value of $\bar{\sigma}_{\chi e} $ above which the CRBDM flux as well as $\frac{dR}{dE_R}(E_R = E_R^{\rm th})$ become negligible.

  The above discussion is a good enough approximation for energy independent cross section consideration and for at most a single scattering scenario.
  However, the above approach is not directly applicable to our energy-dependent scenario (with multiple scattering).
  Thus, we need to modify the analysis a bit. 
  Our modified analysis procedure for the attenuation effect is summarized as follows.
 
\begin{itemize}
\item  We calculate the average energy transfer per scattering in the Earth crust by the CRBDM particles. 
The energy transfer in a single scattering as well as the overall cross section in that scattering are dependent on the kinetic energy of the CRBDM before the scattering occurs.
 
\item  We fix the final energy of the attenuated CRBDM at the detector depth such that it can produce the electron recoil with recoil energy larger than the lower threshold of the detector. 
By estimating the average energy transfer and the kinetic energy of CRBDM before the scattering, we calculate the cross section as well as the mean free path.
 
\item   If the mean free path is less than the detector depth we redo the analysis by considering the kinetic energy before one scattering as the final energy. 
At each step, the mean free path is different and dependent on the kinetic energy of CRBDM before scattering.
 
\item  We repeat the whole process until the combined mean free path overshoots the detector depth, and from the number of required iterations we obtain the number of scatterings before reaching the detector.

\item   At the last iteration, we find the initial kinetic energy of CRBDM before entering the Earth crust. 
Based on the CRBDM flux at the surface, we finally estimate the corresponding flux at the detector depth.
 
\end{itemize}
 
 For different normalized cross section, we estimate the final flux and find out the value at which the CRBDM flux drops to a significantly lower value. 
 Those cross sections cause higher number of interactions before reaching the detector, and are thus unable to produce enough recoil to be detected. 
 We set the upper bound by finding out such a value at which the suppression starts.

\section{Conclusion}
\label{sec:conclusion}

In this work, we examined the cosmic-ray boosted dark matter taking into account the boost due to all the leading components of cosmic rays, namely, electron, proton, and helium nuclei. 
We computed the flux of CRBDM for three benchmark models: the secluded dark photon, $U(1)_{L_e - L_\mu}$, and $U(1)_{B - L}$. 
Knowing the flux, we estimated the event rate expected at the XENONnT experiment, and analyzed the exclusion limits on the coupling parameters of the models. 
Finally, we translated these exclusion limits to the exclusion bounds on the scattering cross section between DM and electron. 
We also devised a semi-analytic approach to obtain the upper limit on the cross section due to the Earth shielding effect which is an approximate but faster method ever presented in this context. 
We found that for the dark-matter mass less than $\mathcal{O}$(keV), the lower exclusion limit with a light mediator is stronger than the limit obtained from the analysis assuming energy independent cross section; the limit with a heavy mediator is stronger than the limit estimated utilizing the energy independent cross section, for most of the DM mass range. 
Of course, this tendency is dependent on the definition of the normalized cross section; in our work, we estimated the cross section with the conventional value of the momentum transfer in a scattering.

For thermally produced DM, the limit is conventionally drawn down to 1 keV DM mass because the thermally produced fermionic DM of mass less than $ \mathcal{O}(1)$ keV would be hot ($v \gg 10^{-3}c$) and cannot be contained within the galactic halo as a structural part of it. 
The mass of the fermionic DM is additionally constrained to be $ \gtrsim \mathcal{O}(100)$ eV by the Pauli exclusion principle~\cite{Tremaine:1979we}; however, the recent study on ultralight fermionic dark matter~\cite{Davoudiasl:2020uig} shows that the constraint can be relaxed by an order of 16.\footnote{For the discussion on the lower bound of bosonic DM mass, see, e.g., Ref.~\cite{Ban:2022jgm}.} 
Thus, although for completeness we consider the light DM down to 1 eV, the validity of fermionic DM of mass less than $ \mathcal{O}(1)$ keV is still debatable. 
Even if we evade the mass constraint following Ref.~\cite{Davoudiasl:2020uig}, the constraints on the cross section corresponding to low mass DM ($\mathcal{O}(1)$ eV) would correspond either to thermally produced hot DM or to non-thermally produced cold DM. 
In both the cases, the constraints on the cross section will depend on the exact value of $\rho_\chi$ which, however, cannot be determined precisely with the current understanding.

\acknowledgments 
We would like to thank Christopher V. Cappiello for useful discussions.
This work was supported by the National Research Foundation of Korea grant funded by the Korea government(MSIT) (NRF-2019R1C1C1005073, RS-2024-00356960).
Hospitality at APCTP during the program ``Dark Matter as a Portal to New Physics'' is kindly acknowledged.

\appendix 

\section{Velocity Distribution}
\label{sec:vel_dist}

\begin{figure*}
 \centering
 \includegraphics[width=0.49 \textwidth]{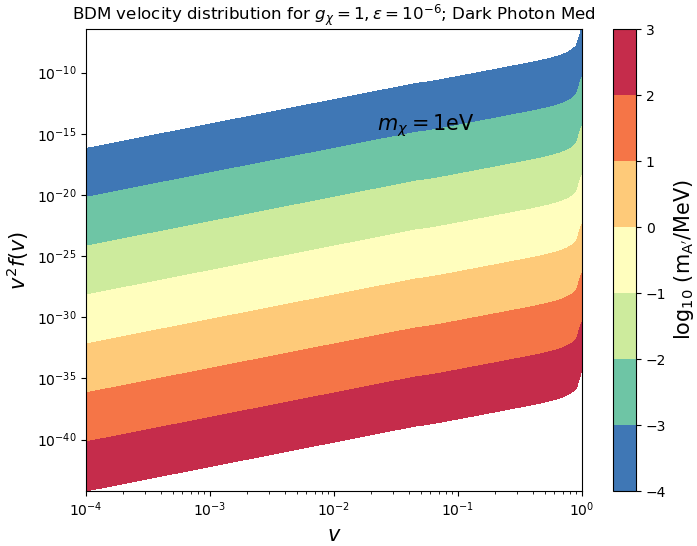}
 \includegraphics[width=0.49 \textwidth]{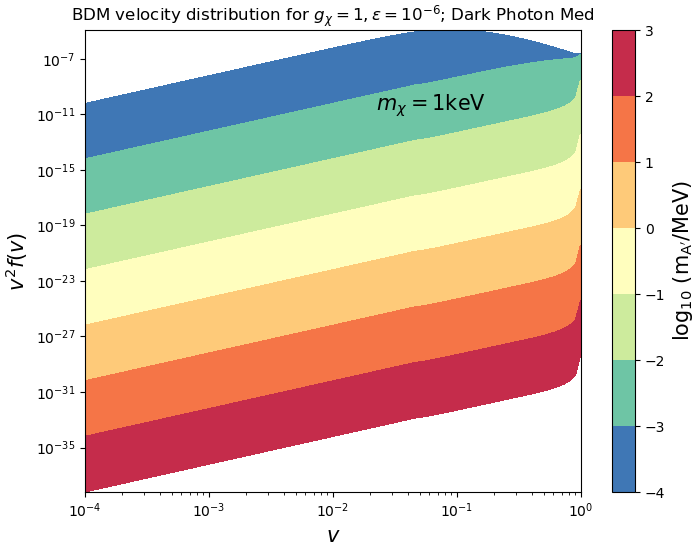}\\
 \includegraphics[width=0.49 \textwidth]{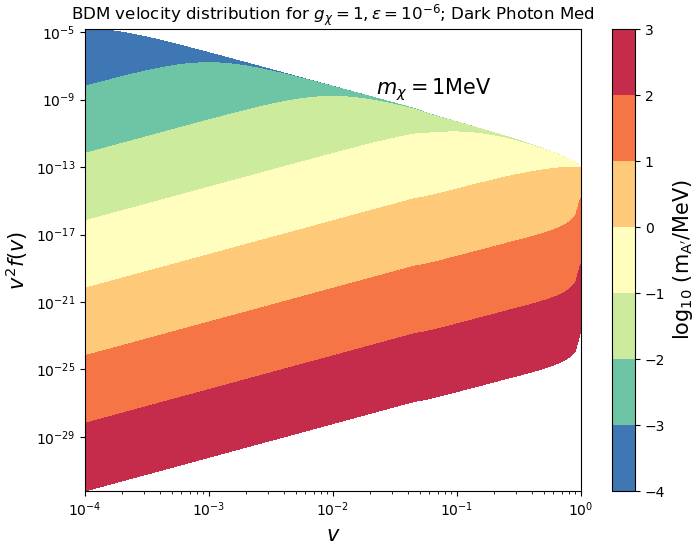}
 \includegraphics[width=0.49 \textwidth]{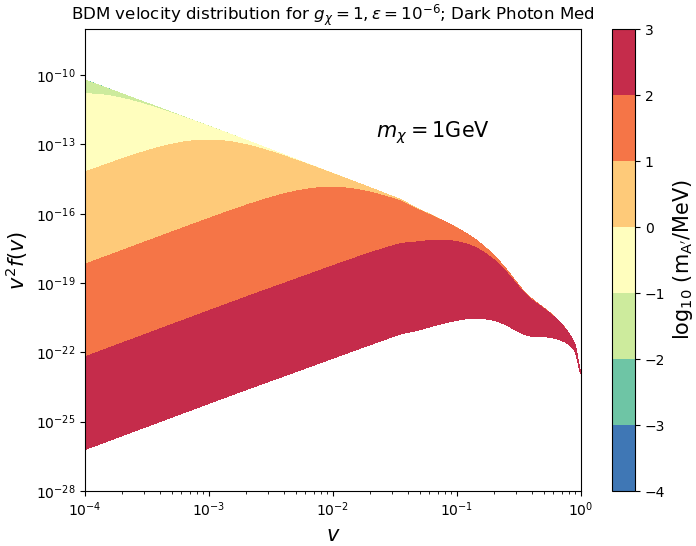}
 \caption{Velocity distribution of CRBDM for $g_\chi = 1, \varepsilon = 10^{-6}$, and $m_\chi = 1~\rm{eV},1~\rm{keV},1~\rm{MeV}, 1~\rm{GeV}$, respectively.}
 \label{fig:vel_dist}
 \end{figure*}

\begin{figure*}
 \centering
 \includegraphics[width=0.49 \textwidth]{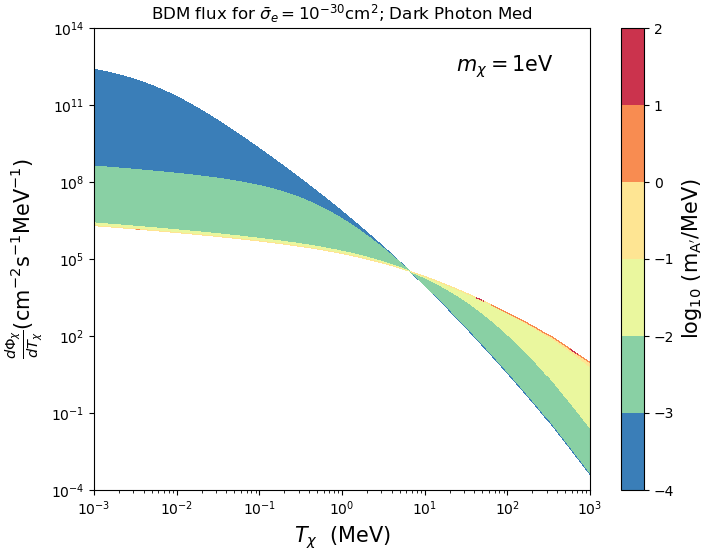}
 \includegraphics[width=0.49 \textwidth]{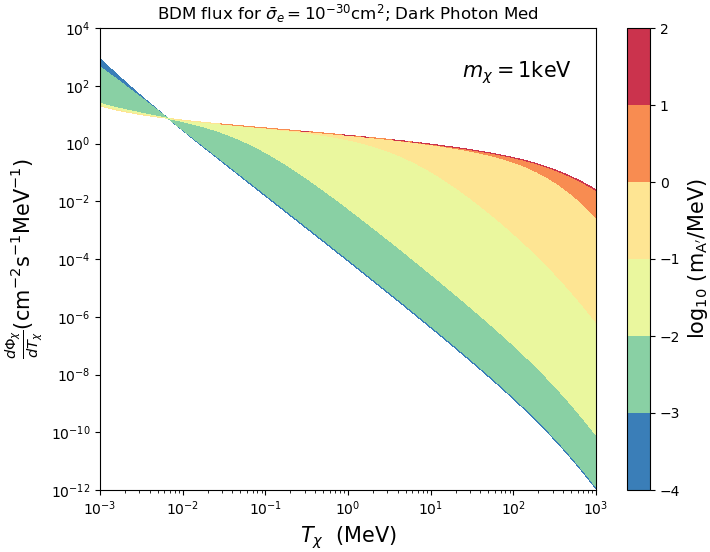}\\
 \includegraphics[width=0.49 \textwidth]{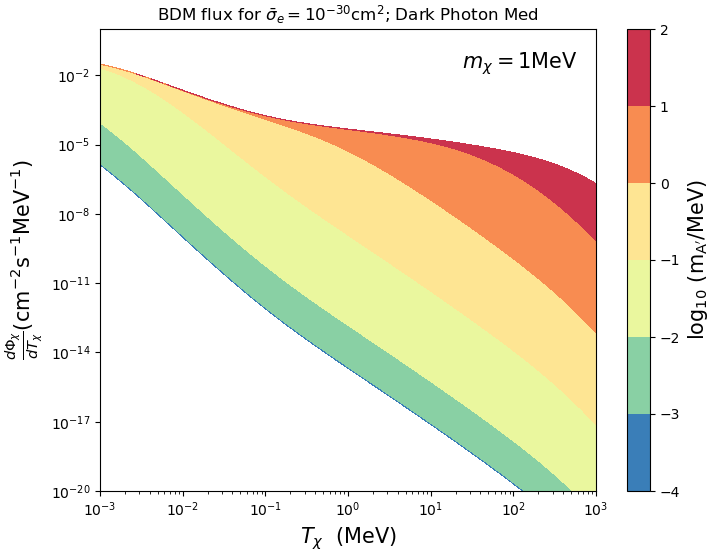}
 \includegraphics[width=0.49 \textwidth]{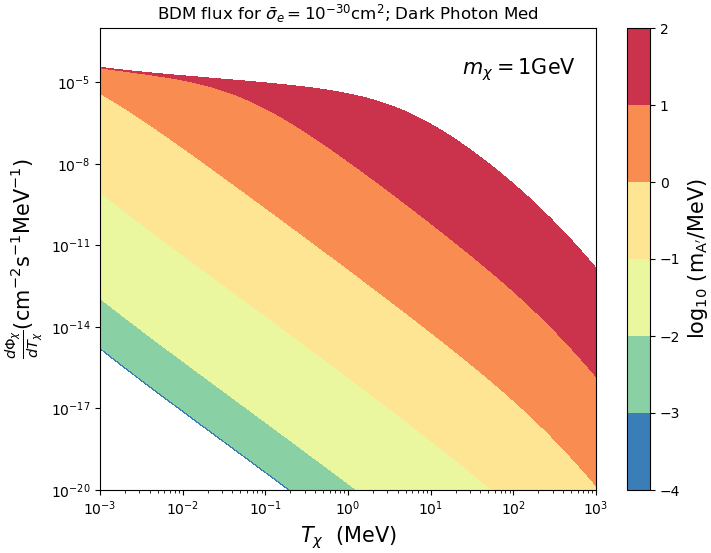} 
 \caption{Cosmic-ray electron boosted DM fluxes for $\bar{\sigma}_e = 10^{-30}~\rm{cm^2}$ and $m_\chi = 1~\rm{eV},1~\rm{keV}, 1~\rm{MeV}, 1~\rm{GeV}$, respectively.}
 \label{fig:flux_eps2}
 \end{figure*}

We show the velocity distribution of the cosmic-ray boosted dark matter in Fig.~\ref{fig:vel_dist} choosing $g_\chi = 1, \varepsilon = 10^{-6}$.  
 To compare with Ref.~\cite{Cao:2020bwd}, we rewrite Eq.~(\ref{eq:diff_cs_p}) considering the interaction only with electron as 
  \bea
 \frac{d \sigma_{\chi e}}{dT_\chi} = \bar{\sigma}_e \frac{\left(\alpha^2 m_e^2 + m_{A'}^2 \right)^2 }{\mu^2_{\chi e}} \frac{2m_\chi \left(me + T_{\rm{CR}} \right)^2 - T_\chi \left\lbrace \left( m_e + m_\chi \right)^2 + 2 m_\chi T_{\rm{CR}}  \right\rbrace + m_\chi T_\chi^2}{4 \left(2 m_e T_{\rm{CR}} + T_{\rm{CR}}^2 \right) \left(2 m_\chi T_\chi + m_{A'}^2 \right)^2}\,,
 \nonumber \\
 \label{eq:diff_cs2}
 \eea
%
 where $\alpha$ is the electromagnetic fine-structure constant and $\mu_{\chi e}$ is the reduced mass of electron and DM. 
 In addition, the normalized DM-electron scattering cross section is given as
 \bea
 \bar{\sigma}_{\chi e} = \frac{\left({g_{eL}^{A'}}^2  + {g_{eR}^{A'}}^2 \right){g_{\chi}^{A'}}^2 \mu^2_{\chi e}}{\pi \left(\alpha^2 m_e^2 + m_{A'}^2 \right)^2 }\,.
 \eea
 Fixing $\bar{\sigma}_{\chi e} = 10^{-30}~\rm{cm^2}$, we obtain the fluxes for CR electron boosted DM as shown in Fig.~\ref{fig:flux_eps2}.

\section{Comparison of BDM fluxes}
\label{sec:comp_BDM_flux}

\begin{figure*}
 \centering
 \includegraphics[width=0.325 \textwidth]{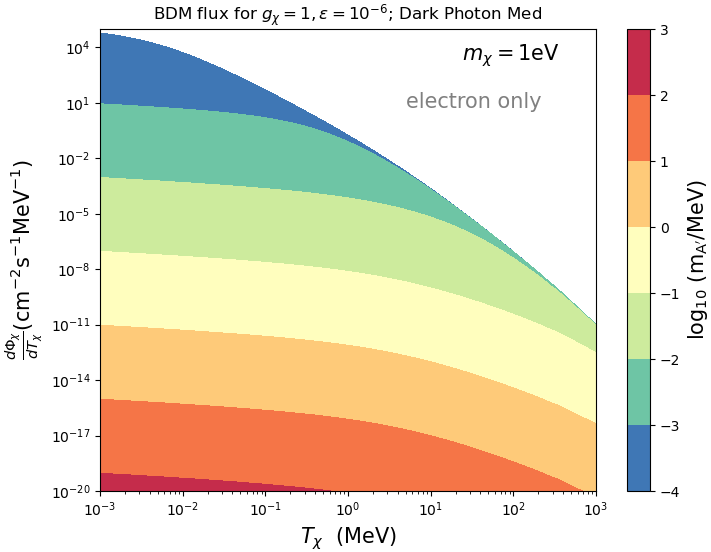}
 \includegraphics[width=0.325 \textwidth]{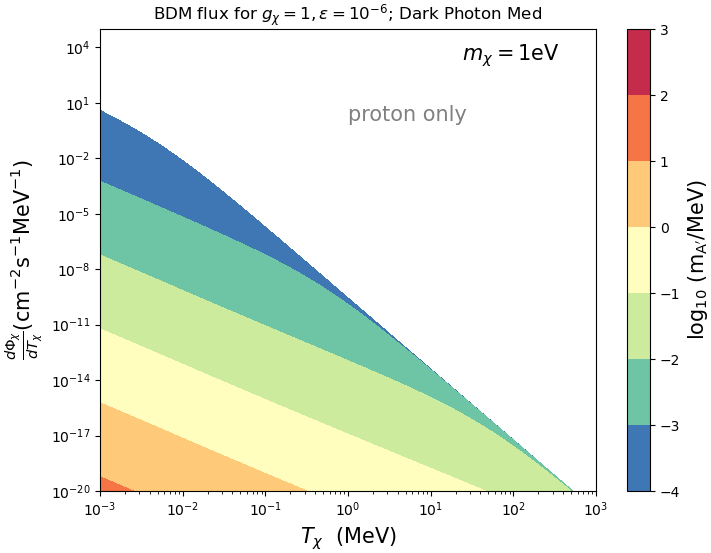}
 \includegraphics[width=0.325 \textwidth]{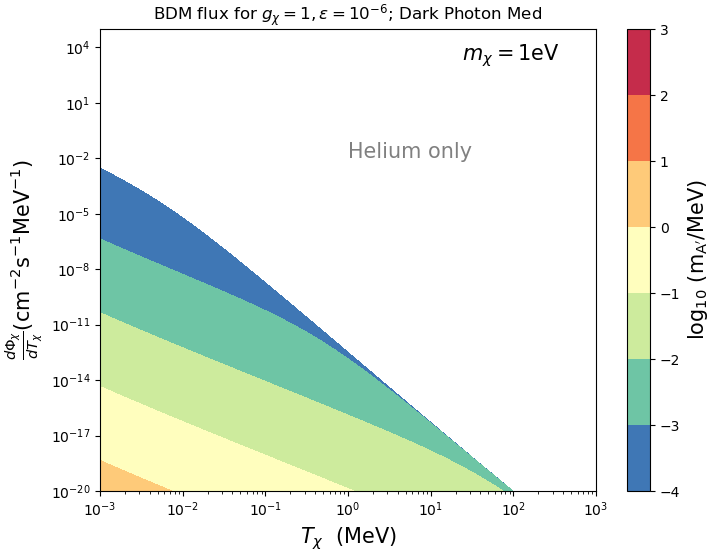}\\
 \includegraphics[width=0.325 \textwidth]{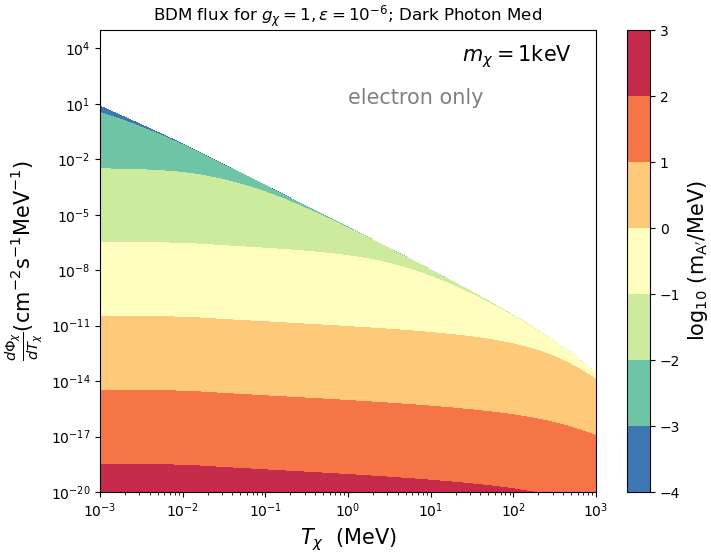}
 \includegraphics[width=0.325 \textwidth]{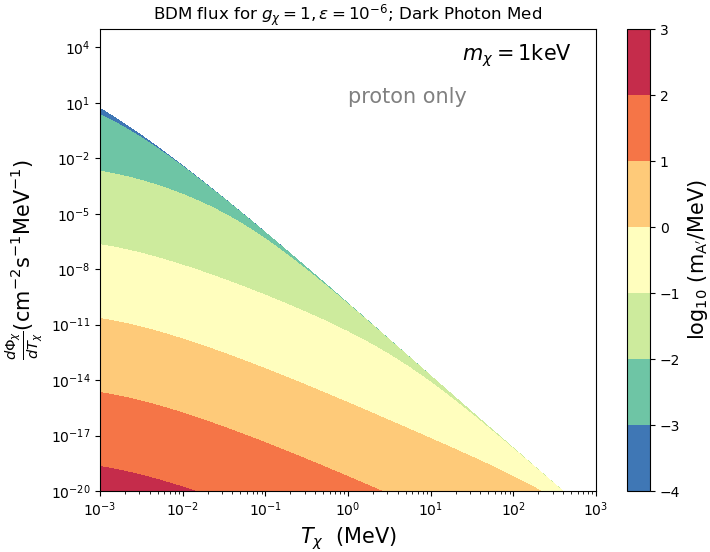}
 \includegraphics[width=0.325 \textwidth]{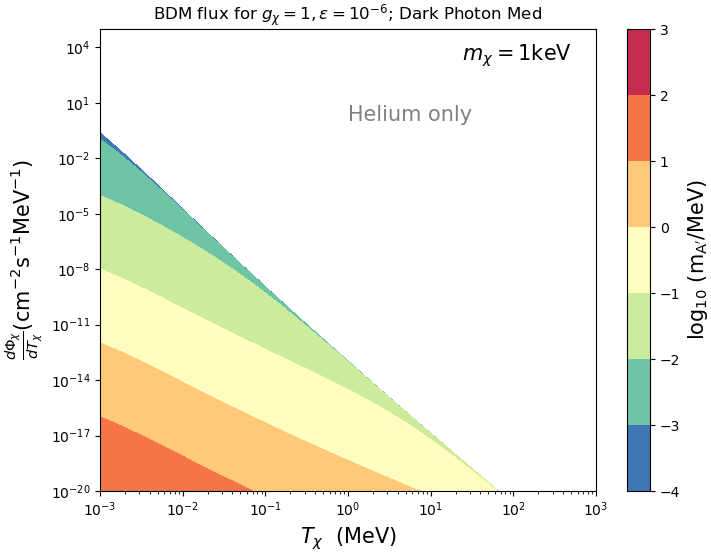}\\
 \includegraphics[width=0.325 \textwidth]{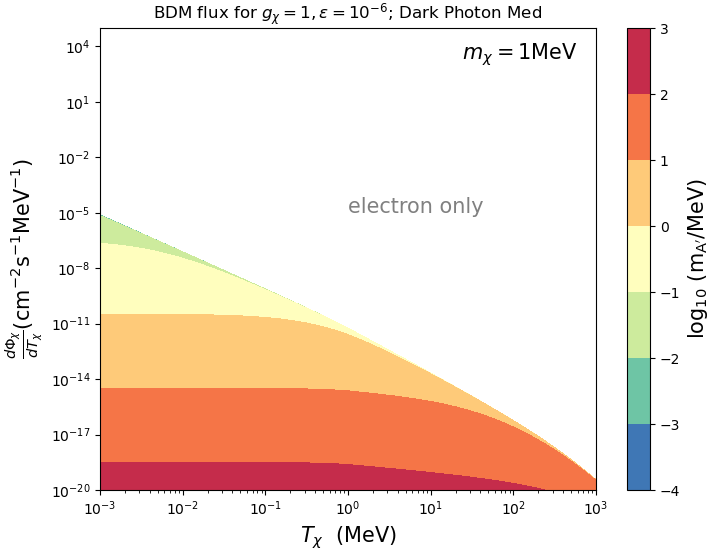}
 \includegraphics[width=0.325 \textwidth]{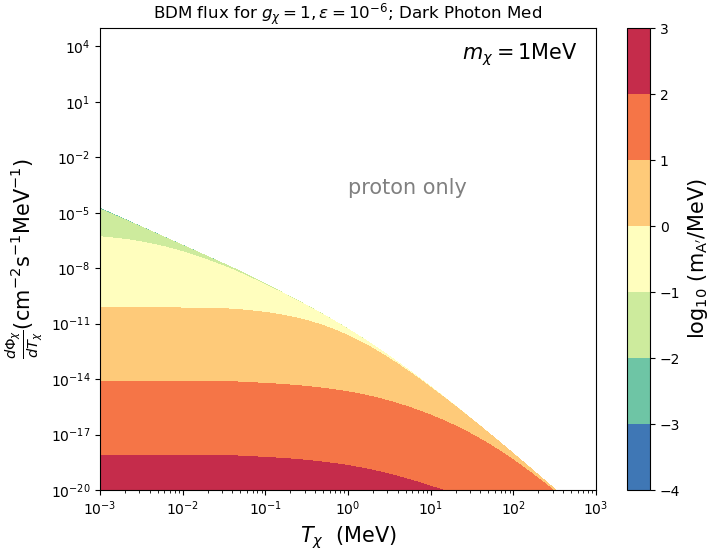}
 \includegraphics[width=0.325 \textwidth]{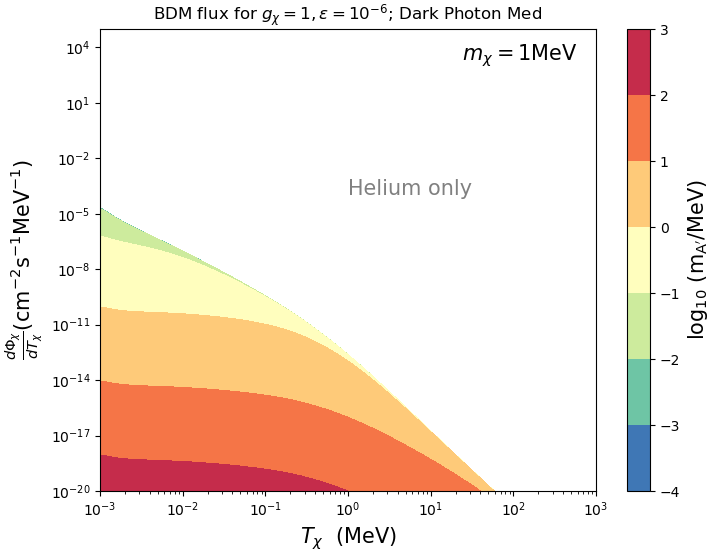}
 \caption{For the secluded dark photon model, cosmic-ray electron (left), proton (middle), and helium (right) boosted DM fluxes for $g_\chi = 1, \varepsilon = 10^{-6}$, and $m_\chi = 1~\rm{eV},1~\rm{keV},1~\rm{MeV}$, respectively.}
 \label{Fig:extra_flux}
  \end{figure*}

\begin{figure*}
 \centering
 \includegraphics[width=0.325 \textwidth]{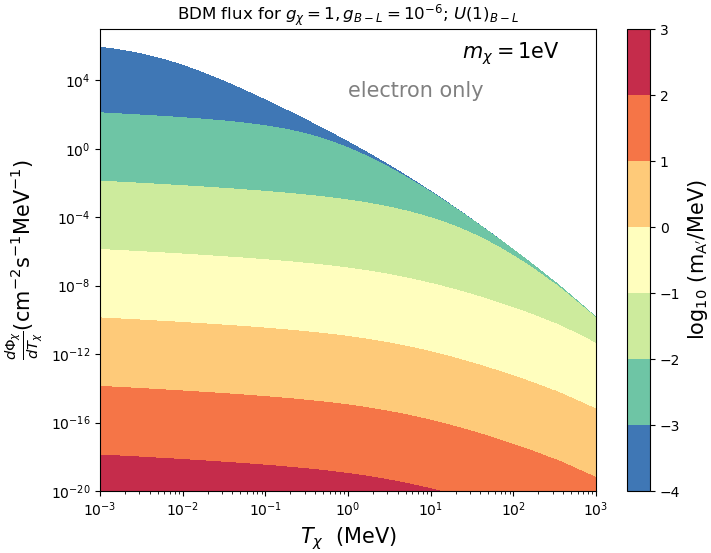}
 \includegraphics[width=0.325 \textwidth]{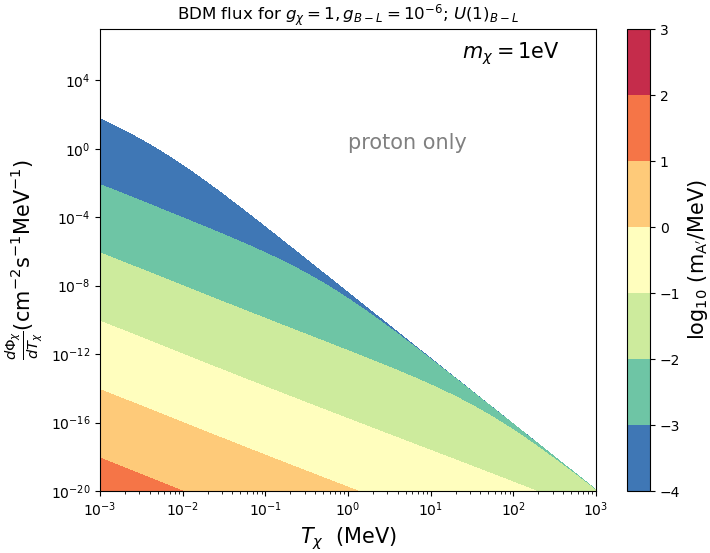}
 \includegraphics[width=0.325 \textwidth]{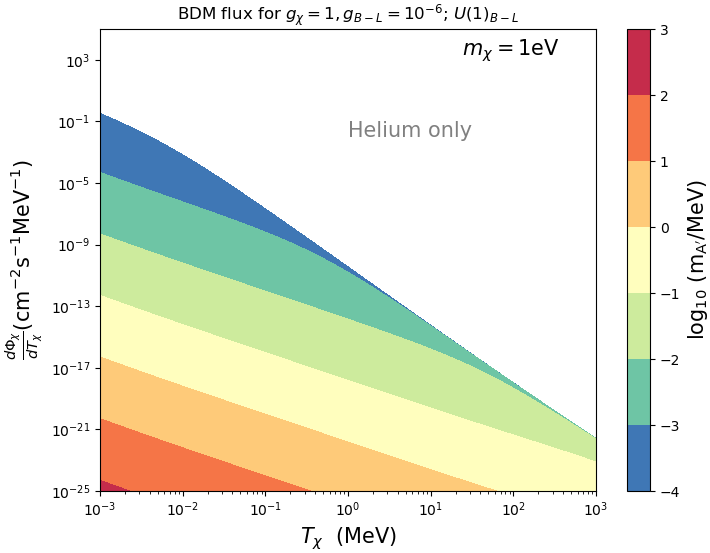}\\
 \includegraphics[width=0.325 \textwidth]{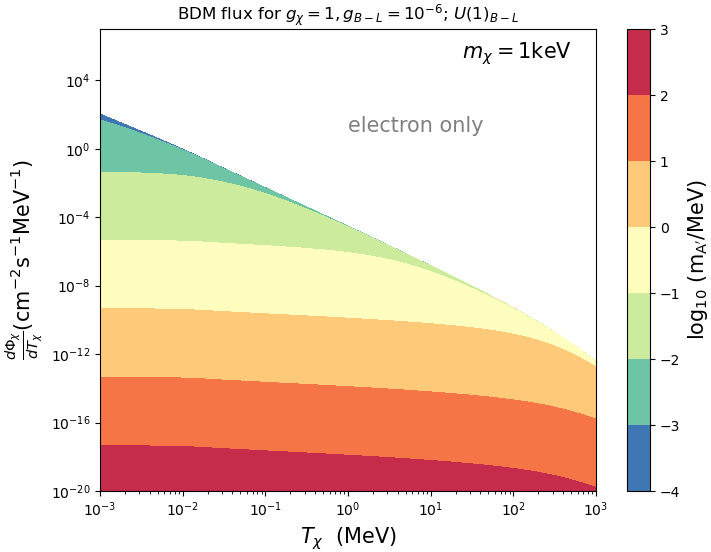}
 \includegraphics[width=0.325 \textwidth]{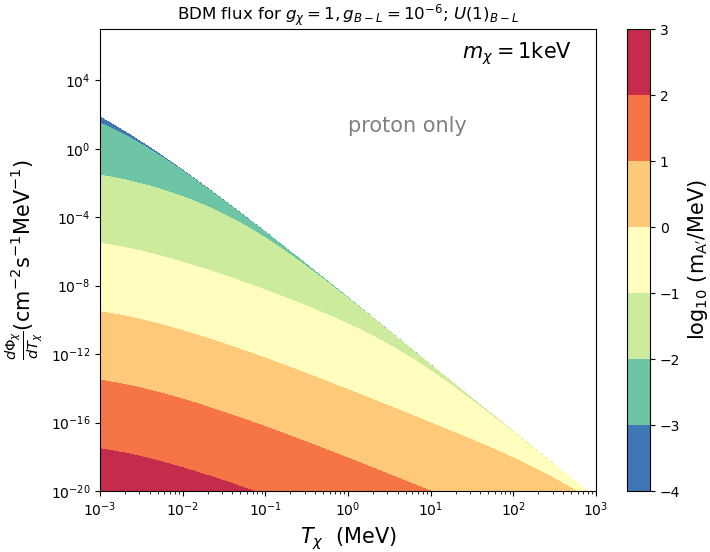}
 \includegraphics[width=0.325 \textwidth]{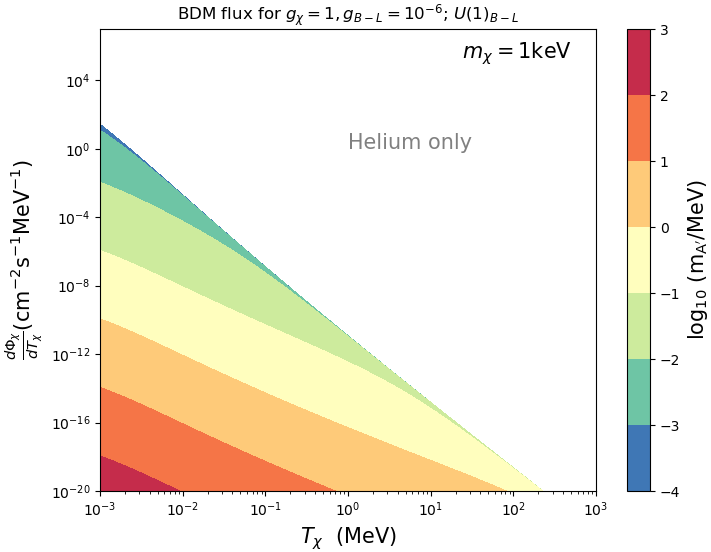}\\
 \includegraphics[width=0.325 \textwidth]{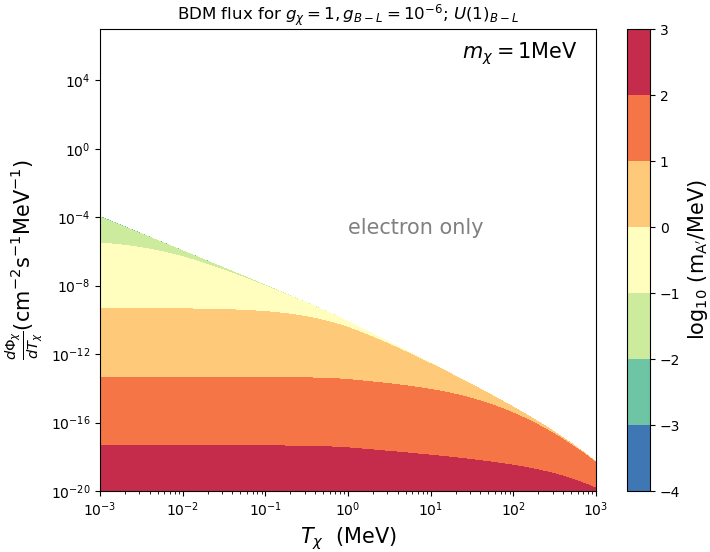}
 \includegraphics[width=0.325 \textwidth]{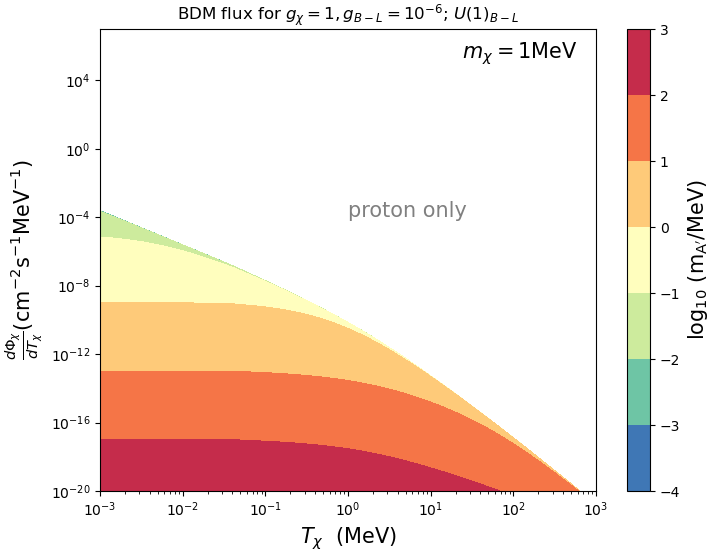}
 \includegraphics[width=0.325 \textwidth]{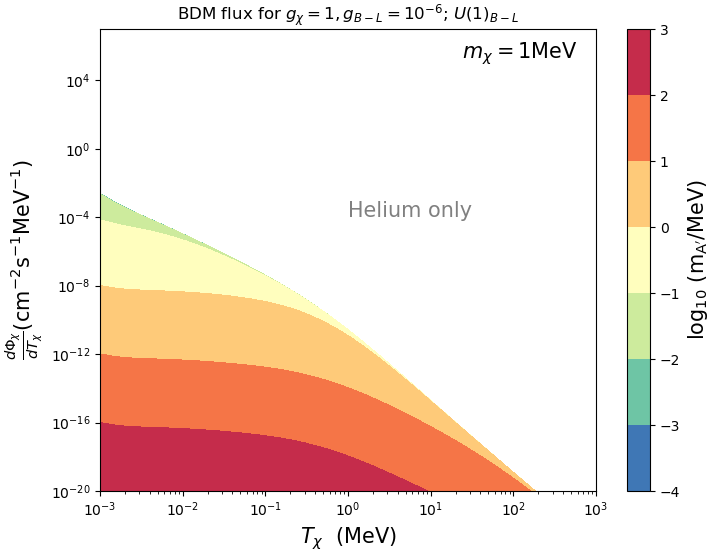}
 \caption{Cosmic-ray electron (left), proton (middle), and helium (right) boosted DM fluxes for the $U(1)_{B-L}$ model.
 Others are the same as Fig.~\ref{Fig:extra_flux}.}
 \label{Fig:extra_flux2}
  \end{figure*}

\begin{figure*}
 \centering
 \includegraphics[width=0.325 \textwidth]{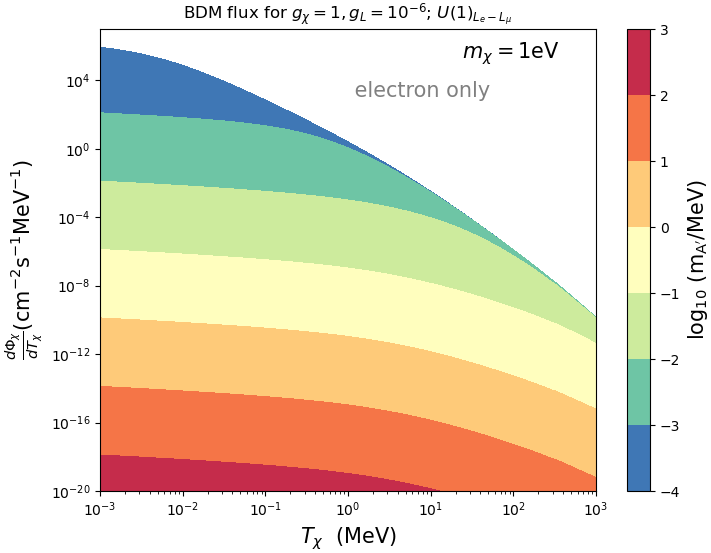}
 \includegraphics[width=0.325 \textwidth]{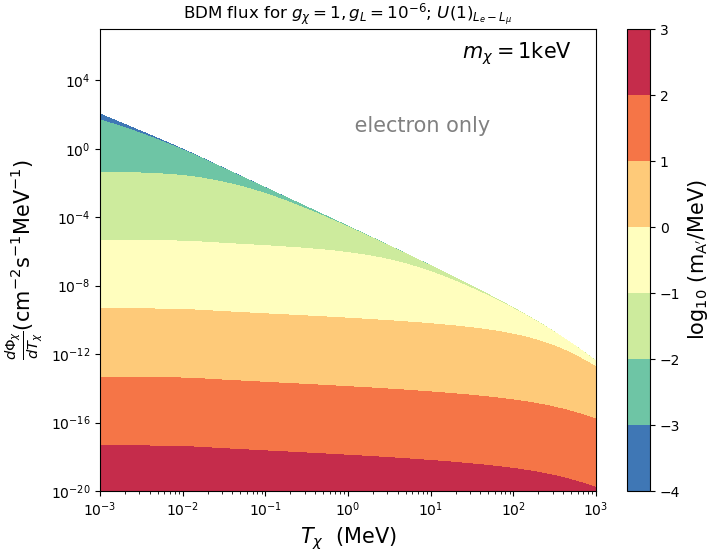}
 \includegraphics[width=0.325 \textwidth]{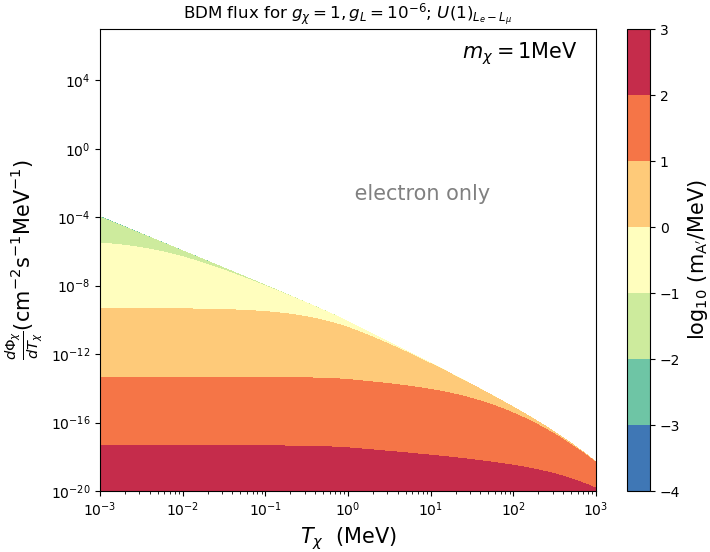}
 \caption{Cosmic-ray electron boosted DM fluxes for the $U(1)_{L_e - L_\mu}$ model.
 Others are the same as Fig.~\ref{Fig:extra_flux}.}
 \label{Fig:extra_flux3}
  \end{figure*}

In Fig.~\ref{Fig:extra_flux}, we separately show the BDM fluxes due to each cosmic-ray species, i.e., electron, proton, and helium for comparison, based on the secluded dark photon model. 
Similar plots for the $U(1)_{B-L}$ model and the $L_e - L_\mu$ model are depicted in Fig.~\ref{Fig:extra_flux2} and Fig.~\ref{Fig:extra_flux3}, respectively. 
Note that only cosmic-ray electron boosted DM flux exists for the $L_e - L_\mu$ model.

\section{Calculation of amplitude and cross section}
\label{sec:cs_est}

 Here we present a simple analytical form for a quick estimation of cross section for the secluded dark photon model. 
 We follow the relativistic kinematics of two-body scattering process while the second particle is at rest. 
 We obtain Eq.~(\ref{Eq:T-chi-max}) and thereafter the differential cross section with respect to the kinetic energy~\cite{Cao:2020bwd,Cho:2020mnc}
 \bea
\frac{d\cos \theta}{dT_\chi} &=& -\frac{2}{T_\chi^{\rm max}}\,, \nonumber \\
\frac{d\sigma_{\chi e}}{dT_\chi} &=& \frac{d\sigma_{\chi e}}{d\Omega} \cdot \frac{d\Omega}{dT_\chi} = \frac{\overline{\abs{\mathcal{M}}^2}}{16 \pi s} \frac{1}{T_\chi^{\rm max}}\,.
\label{Eq:amp_to_diff_cs}
 \eea
 The corresponding Mandelstam variables are given by
 \bea
 s &=& \left( m_e + m_\chi \right)^2 + 2 T_e m_\chi\,, \nonumber \\
 t &=& - 2 m_\chi T_\chi\,, \nonumber \\
 u &=& 2 \left(m_e^2 + m_\chi^2 \right) -s -t\,.
 \eea
 The leading contribution of the DM-SM interaction comes from the dark photon mediated channel. 
 However, to take into account the small contributions of the SM $Z$ boson as well, we follow~\cite{Cho:2020mnc}. 
 The complete compact forms are given below.

\subsection{DM-electron scattering}
 
 For DM-electron interaction, the spin-averaged amplitude square is given by
  \bea
 \overline{\abs{\mathcal{M}}^2} = \frac{2 g_\chi^2}{1 - \varepsilon^2} \Biggl[\left(\frac{s_\xi g_C}{t-m_Z^2} - \frac{c_\xi g_{Cd}}{t-m_{A^{\prime}}^2}\right)^2 ~ A(m_\chi, m_e) + \nonumber \\ \left(\frac{s_\xi g_A}{t-m_Z^2} - \frac{c_\xi g_{Ad}}{t-m_{A^{\prime}}^2}\right)^2 ~ B(m_\chi, m_e)  \Biggr]\,,
 \eea
 which we have to plug into Eq.~(\ref{Eq:amp_to_diff_cs}) to estimate the differential cross section.
 Here,  
 \bea
 g_C &=& \frac{e}{4} \left[c_\xi \left(3 \tan \theta_W - \cot \theta_W \right) + \frac{3 s_\xi t_\varepsilon}{c_W} \right]\,, \nonumber \\
  g_{Cd} &=& \frac{e}{4} \left[s_\xi \left(3 \tan \theta_W - \cot \theta_W \right) - \frac{3 c_\xi t_\varepsilon}{c_W} \right]\,, \nonumber \\
  g_A &=& \frac{e}{4 c_W} \left[ \frac{c_\xi}{s_W} + s_\xi t_\varepsilon \right]\,, \nonumber \\
 g_{Ad} &=& \frac{e}{4 c_W} \left[ \frac{s_\xi}{s_W} - c_\xi t_\varepsilon \right]\,, 
 \eea
 and assuming $M^2 = \left(m_\chi^2 + m_i^2 \right)$ the functions $A, B$ are given as follows 
 \bea
 A(m_\chi, m_i) &=& 2 t M^2 + \left(s-M^2 \right)^2 + \left(u-M^2 \right)^2\,, \nonumber \\
 B(m_\chi, m_i) &=&  \left(s-M^2 \right)^2 + \left(u-M^2 \right)^2 + 2 t \left(m_\chi^2 - m_i^2 \right) - 8 m_\chi^2 m_i^2 \,.
 \eea

\subsection{DM-nucleus scattering}
  For DM-nucleus interaction, the spin-averaged amplitude square is given by
 \bea
 \overline{\abs{\mathcal{M}}^2} = \frac{2 g_\chi^2}{1 - \varepsilon^2} \Biggl[\left(\frac{s_\xi g_{NC}}{t-m_Z^2} - \frac{c_\xi g_{NCd}}{t-m_{A^{\prime}}^2}\right)^2 ~ A(m_\chi, m_N) + \nonumber \\ \left(\frac{s_\xi g_{NA}}{t-m_Z^2} - \frac{c_\xi g_{NAd}}{t-m_{A^{\prime}}^2}\right)^2 ~ B(m_\chi, m_N)  \Biggr]\,, 
 \eea
 where assuming $M^2 = \left(m_\chi^2 + m_N^2 \right)$ the functions $A, B$ are given as
 \bea
 A(m_\chi, m_N) &=& 2 t M^2 + \left(s-M^2 \right)^2 + \left(u-M^2 \right)^2\,, \nonumber \\
 B(m_\chi, m_N) &=&  \left(s-M^2 \right)^2 + \left(u-M^2 \right)^2 + 2 t \left(m_\chi^2 - m_N^2 \right) - 8 m_\chi^2 m_N^2 \,.
 \eea
  Again we have to plug this amplitude into Eq.~(\ref{Eq:amp_to_diff_cs}) to estimate the differential cross section.
 For the nucleus with atomic number $Z$ and mass number $A$, we have 
 \bea
 g_{NC} &=& Z g_C + (A - Z) g_A\,, \nonumber \\
  g_{NCd} &=& Z g_{Cd} + (A - Z) g_{Ad}\,, \nonumber \\
  g_{NA} &=& (2 Z - A) g_{A}\,, \nonumber \\
 g_{NAd} &=& (2 Z - A) g_{Ad} \,.
 \eea

\bibliographystyle{JHEP}
\bibliography{ref}

\providecommand{\href}[2]{#2}\begingroup\raggedright\begin{thebibliography}{100}

\bibitem{Planck:2015fie}
{\scshape Planck} collaboration, \emph{{Planck 2015 results. XIII. Cosmological
  parameters}},
  \href{https://doi.org/10.1051/0004-6361/201525830}{\emph{Astron. Astrophys.}
  {\bfseries 594} (2016) A13}
  [\href{https://arxiv.org/abs/1502.01589}{{\ttfamily 1502.01589}}].

\bibitem{Bertone:2004pz}
G.~Bertone, D.~Hooper and J.~Silk, \emph{{Particle dark matter: Evidence,
  candidates and constraints}},
  \href{https://doi.org/10.1016/j.physrep.2004.08.031}{\emph{Phys. Rept.}
  {\bfseries 405} (2005) 279}
  [\href{https://arxiv.org/abs/hep-ph/0404175}{{\ttfamily hep-ph/0404175}}].

\bibitem{Goodman:1984dc}
M.W.~Goodman and E.~Witten, \emph{{Detectability of Certain Dark Matter
  Candidates}}, \href{https://doi.org/10.1103/PhysRevD.31.3059}{\emph{Phys.
  Rev. D} {\bfseries 31} (1985) 3059}.

\bibitem{DAMA:2008jlt}
{\scshape DAMA} collaboration, \emph{{First results from DAMA/LIBRA and the
  combined results with DAMA/NaI}},
  \href{https://doi.org/10.1140/epjc/s10052-008-0662-y}{\emph{Eur. Phys. J. C}
  {\bfseries 56} (2008) 333} [\href{https://arxiv.org/abs/0804.2741}{{\ttfamily
  0804.2741}}].

\bibitem{CRESST:2015txj}
{\scshape CRESST} collaboration, \emph{{Results on light dark matter particles
  with a low-threshold CRESST-II detector}},
  \href{https://doi.org/10.1140/epjc/s10052-016-3877-3}{\emph{Eur. Phys. J. C}
  {\bfseries 76} (2016) 25} [\href{https://arxiv.org/abs/1509.01515}{{\ttfamily
  1509.01515}}].

\bibitem{DarkSide:2018ppu}
{\scshape DarkSide} collaboration, \emph{{Constraints on Sub-GeV
  Dark-Matter\textendash{}Electron Scattering from the DarkSide-50
  Experiment}},
  \href{https://doi.org/10.1103/PhysRevLett.121.111303}{\emph{Phys. Rev. Lett.}
  {\bfseries 121} (2018) 111303}
  [\href{https://arxiv.org/abs/1802.06998}{{\ttfamily 1802.06998}}].

\bibitem{SuperCDMS:2018mne}
{\scshape SuperCDMS} collaboration, \emph{{First Dark Matter Constraints from a
  SuperCDMS Single-Charge Sensitive Detector}},
  \href{https://doi.org/10.1103/PhysRevLett.121.051301}{\emph{Phys. Rev. Lett.}
  {\bfseries 121} (2018) 051301}
  [\href{https://arxiv.org/abs/1804.10697}{{\ttfamily 1804.10697}}].

\bibitem{XENONCollaboration:2023orw}
{\scshape (XENON Collaboration)\textdagger{}\textdagger{}, XENON}
  collaboration, \emph{{First Dark Matter Search with Nuclear Recoils from the
  XENONnT Experiment}},
  \href{https://doi.org/10.1103/PhysRevLett.131.041003}{\emph{Phys. Rev. Lett.}
  {\bfseries 131} (2023) 041003}
  [\href{https://arxiv.org/abs/2303.14729}{{\ttfamily 2303.14729}}].

\bibitem{LUX:2022vee}
{\scshape LUX} collaboration, \emph{{Fast and flexible analysis of direct dark
  matter search data with machine learning}},
  \href{https://doi.org/10.1103/PhysRevD.106.072009}{\emph{Phys. Rev. D}
  {\bfseries 106} (2022) 072009}
  [\href{https://arxiv.org/abs/2201.05734}{{\ttfamily 2201.05734}}].

\bibitem{PandaX-4T:2021bab}
{\scshape PandaX-4T} collaboration, \emph{{Dark Matter Search Results from the
  PandaX-4T Commissioning Run}},
  \href{https://doi.org/10.1103/PhysRevLett.127.261802}{\emph{Phys. Rev. Lett.}
  {\bfseries 127} (2021) 261802}
  [\href{https://arxiv.org/abs/2107.13438}{{\ttfamily 2107.13438}}].

\bibitem{LUX-ZEPLIN:2022xrq}
{\scshape LUX-ZEPLIN} collaboration, \emph{{First Dark Matter Search Results
  from the LUX-ZEPLIN (LZ) Experiment}},
  \href{https://doi.org/10.1103/PhysRevLett.131.041002}{\emph{Phys. Rev. Lett.}
  {\bfseries 131} (2023) 041002}
  [\href{https://arxiv.org/abs/2207.03764}{{\ttfamily 2207.03764}}].

\bibitem{XENON:2022ltv}
{\scshape XENON} collaboration, \emph{{Search for New Physics in Electronic
  Recoil Data from XENONnT}},
  \href{https://doi.org/10.1103/PhysRevLett.129.161805}{\emph{Phys. Rev. Lett.}
  {\bfseries 129} (2022) 161805}
  [\href{https://arxiv.org/abs/2207.11330}{{\ttfamily 2207.11330}}].

\bibitem{DAMIC-M:2023gxo}
{\scshape DAMIC-M} collaboration, \emph{{First Constraints from DAMIC-M on
  Sub-GeV Dark-Matter Particles Interacting with Electrons}},
  \href{https://doi.org/10.1103/PhysRevLett.130.171003}{\emph{Phys. Rev. Lett.}
  {\bfseries 130} (2023) 171003}
  [\href{https://arxiv.org/abs/2302.02372}{{\ttfamily 2302.02372}}].

\bibitem{SuperCDMS:2020ymb}
{\scshape SuperCDMS} collaboration, \emph{{Constraints on low-mass, relic dark
  matter candidates from a surface-operated SuperCDMS single-charge sensitive
  detector}}, \href{https://doi.org/10.1103/PhysRevD.102.091101}{\emph{Phys.
  Rev. D} {\bfseries 102} (2020) 091101}
  [\href{https://arxiv.org/abs/2005.14067}{{\ttfamily 2005.14067}}].

\bibitem{SENSEI:2020dpa}
{\scshape SENSEI} collaboration, \emph{{SENSEI: Direct-Detection Results on
  sub-GeV Dark Matter from a New Skipper-CCD}},
  \href{https://doi.org/10.1103/PhysRevLett.125.171802}{\emph{Phys. Rev. Lett.}
  {\bfseries 125} (2020) 171802}
  [\href{https://arxiv.org/abs/2004.11378}{{\ttfamily 2004.11378}}].

\bibitem{EDELWEISS:2020fxc}
{\scshape EDELWEISS} collaboration, \emph{{First germanium-based constraints on
  sub-MeV Dark Matter with the EDELWEISS experiment}},
  \href{https://doi.org/10.1103/PhysRevLett.125.141301}{\emph{Phys. Rev. Lett.}
  {\bfseries 125} (2020) 141301}
  [\href{https://arxiv.org/abs/2003.01046}{{\ttfamily 2003.01046}}].

\bibitem{CRESST:2019jnq}
{\scshape CRESST} collaboration, \emph{{First results from the CRESST-III
  low-mass dark matter program}},
  \href{https://doi.org/10.1103/PhysRevD.100.102002}{\emph{Phys. Rev. D}
  {\bfseries 100} (2019) 102002}
  [\href{https://arxiv.org/abs/1904.00498}{{\ttfamily 1904.00498}}].

\bibitem{Hochberg:2015pha}
Y.~Hochberg, Y.~Zhao and K.M.~Zurek, \emph{{Superconducting Detectors for
  Superlight Dark Matter}},
  \href{https://doi.org/10.1103/PhysRevLett.116.011301}{\emph{Phys. Rev. Lett.}
  {\bfseries 116} (2016) 011301}
  [\href{https://arxiv.org/abs/1504.07237}{{\ttfamily 1504.07237}}].

\bibitem{Schutz:2016tid}
K.~Schutz and K.M.~Zurek, \emph{{Detectability of Light Dark Matter with
  Superfluid Helium}},
  \href{https://doi.org/10.1103/PhysRevLett.117.121302}{\emph{Phys. Rev. Lett.}
  {\bfseries 117} (2016) 121302}
  [\href{https://arxiv.org/abs/1604.08206}{{\ttfamily 1604.08206}}].

\bibitem{Hochberg:2017wce}
Y.~Hochberg, Y.~Kahn, M.~Lisanti, K.M.~Zurek, A.G.~Grushin, R.~Ilan et~al.,
  \emph{{Detection of sub-MeV Dark Matter with Three-Dimensional Dirac
  Materials}}, \href{https://doi.org/10.1103/PhysRevD.97.015004}{\emph{Phys.
  Rev. D} {\bfseries 97} (2018) 015004}
  [\href{https://arxiv.org/abs/1708.08929}{{\ttfamily 1708.08929}}].

\bibitem{Knapen:2017ekk}
S.~Knapen, T.~Lin, M.~Pyle and K.M.~Zurek, \emph{{Detection of Light Dark
  Matter With Optical Phonons in Polar Materials}},
  \href{https://doi.org/10.1016/j.physletb.2018.08.064}{\emph{Phys. Lett. B}
  {\bfseries 785} (2018) 386}
  [\href{https://arxiv.org/abs/1712.06598}{{\ttfamily 1712.06598}}].

\bibitem{Hochberg:2019cyy}
Y.~Hochberg, I.~Charaev, S.-W.~Nam, V.~Verma, M.~Colangelo and K.K.~Berggren,
  \emph{{Detecting Sub-GeV Dark Matter with Superconducting Nanowires}},
  \href{https://doi.org/10.1103/PhysRevLett.123.151802}{\emph{Phys. Rev. Lett.}
  {\bfseries 123} (2019) 151802}
  [\href{https://arxiv.org/abs/1903.05101}{{\ttfamily 1903.05101}}].

\bibitem{Kim:2020bwm}
D.~Kim, J.-C.~Park, K.C.~Fong and G.-H.~Lee, \emph{{Detection of Super-light
  Dark Matter Using Graphene Sensor}},
  \href{https://arxiv.org/abs/2002.07821}{{\ttfamily 2002.07821}}.

\bibitem{Belanger:2011ww}
G.~Belanger and J.-C.~Park, \emph{{Assisted freeze-out}},
  \href{https://doi.org/10.1088/1475-7516/2012/03/038}{\emph{JCAP} {\bfseries
  03} (2012) 038} [\href{https://arxiv.org/abs/1112.4491}{{\ttfamily
  1112.4491}}].

\bibitem{Agashe:2014yua}
K.~Agashe, Y.~Cui, L.~Necib and J.~Thaler, \emph{{(In)direct Detection of
  Boosted Dark Matter}},
  \href{https://doi.org/10.1088/1475-7516/2014/10/062}{\emph{JCAP} {\bfseries
  10} (2014) 062} [\href{https://arxiv.org/abs/1405.7370}{{\ttfamily
  1405.7370}}].

\bibitem{Berger:2014sqa}
J.~Berger, Y.~Cui and Y.~Zhao, \emph{{Detecting Boosted Dark Matter from the
  Sun with Large Volume Neutrino Detectors}},
  \href{https://doi.org/10.1088/1475-7516/2015/02/005}{\emph{JCAP} {\bfseries
  02} (2015) 005} [\href{https://arxiv.org/abs/1410.2246}{{\ttfamily
  1410.2246}}].

\bibitem{Kong:2014mia}
K.~Kong, G.~Mohlabeng and J.-C.~Park, \emph{{Boosted dark matter signals
  uplifted with self-interaction}},
  \href{https://doi.org/10.1016/j.physletb.2015.02.057}{\emph{Phys. Lett. B}
  {\bfseries 743} (2015) 256}
  [\href{https://arxiv.org/abs/1411.6632}{{\ttfamily 1411.6632}}].

\bibitem{Cherry:2015oca}
J.F.~Cherry, M.T.~Frandsen and I.M.~Shoemaker, \emph{{Direct Detection
  Phenomenology in Models Where the Products of Dark Matter Annihilation
  Interact with Nuclei}},
  \href{https://doi.org/10.1103/PhysRevLett.114.231303}{\emph{Phys. Rev. Lett.}
  {\bfseries 114} (2015) 231303}
  [\href{https://arxiv.org/abs/1501.03166}{{\ttfamily 1501.03166}}].

\bibitem{Necib:2016aez}
L.~Necib, J.~Moon, T.~Wongjirad and J.M.~Conrad, \emph{{Boosted Dark Matter at
  Neutrino Experiments}},
  \href{https://doi.org/10.1103/PhysRevD.95.075018}{\emph{Phys. Rev. D}
  {\bfseries 95} (2017) 075018}
  [\href{https://arxiv.org/abs/1610.03486}{{\ttfamily 1610.03486}}].

\bibitem{Alhazmi:2016qcs}
H.~Alhazmi, K.~Kong, G.~Mohlabeng and J.-C.~Park, \emph{{Boosted Dark Matter at
  the Deep Underground Neutrino Experiment}},
  \href{https://doi.org/10.1007/JHEP04(2017)158}{\emph{JHEP} {\bfseries 04}
  (2017) 158} [\href{https://arxiv.org/abs/1611.09866}{{\ttfamily
  1611.09866}}].

\bibitem{Kim:2016zjx}
D.~Kim, J.-C.~Park and S.~Shin, \emph{{Dark Matter
  \textquotedblleft{}Collider\textquotedblright{} from Inelastic Boosted Dark
  Matter}}, \href{https://doi.org/10.1103/PhysRevLett.119.161801}{\emph{Phys.
  Rev. Lett.} {\bfseries 119} (2017) 161801}
  [\href{https://arxiv.org/abs/1612.06867}{{\ttfamily 1612.06867}}].

\bibitem{Giudice:2017zke}
G.F.~Giudice, D.~Kim, J.-C.~Park and S.~Shin, \emph{{Inelastic Boosted Dark
  Matter at Direct Detection Experiments}},
  \href{https://doi.org/10.1016/j.physletb.2018.03.043}{\emph{Phys. Lett. B}
  {\bfseries 780} (2018) 543}
  [\href{https://arxiv.org/abs/1712.07126}{{\ttfamily 1712.07126}}].

\bibitem{Chatterjee:2018mej}
A.~Chatterjee, A.~De~Roeck, D.~Kim, Z.G.~Moghaddam, J.-C.~Park, S.~Shin et~al.,
  \emph{{Searching for boosted dark matter at ProtoDUNE}},
  \href{https://doi.org/10.1103/PhysRevD.98.075027}{\emph{Phys. Rev. D}
  {\bfseries 98} (2018) 075027}
  [\href{https://arxiv.org/abs/1803.03264}{{\ttfamily 1803.03264}}].

\bibitem{Kim:2018veo}
D.~Kim, K.~Kong, J.-C.~Park and S.~Shin, \emph{{Boosted Dark Matter Quarrying
  at Surface Neutrino Detectors}},
  \href{https://doi.org/10.1007/JHEP08(2018)155}{\emph{JHEP} {\bfseries 08}
  (2018) 155} [\href{https://arxiv.org/abs/1804.07302}{{\ttfamily
  1804.07302}}].

\bibitem{Aoki:2018gjf}
M.~Aoki and T.~Toma, \emph{{Boosted Self-interacting Dark Matter in a
  Multi-component Dark Matter Model}},
  \href{https://doi.org/10.1088/1475-7516/2018/10/020}{\emph{JCAP} {\bfseries
  10} (2018) 020} [\href{https://arxiv.org/abs/1806.09154}{{\ttfamily
  1806.09154}}].

\bibitem{Kim:2019had}
D.~Kim, J.-C.~Park and S.~Shin, \emph{{Searching for boosted dark matter via
  dark-photon bremsstrahlung}},
  \href{https://doi.org/10.1103/PhysRevD.100.035033}{\emph{Phys. Rev. D}
  {\bfseries 100} (2019) 035033}
  [\href{https://arxiv.org/abs/1903.05087}{{\ttfamily 1903.05087}}].

\bibitem{Kim:2020ipj}
D.~Kim, P.A.N.~Machado, J.-C.~Park and S.~Shin, \emph{{Optimizing Energetic
  Light Dark Matter Searches in Dark Matter and Neutrino Experiments}},
  \href{https://doi.org/10.1007/JHEP07(2020)057}{\emph{JHEP} {\bfseries 07}
  (2020) 057} [\href{https://arxiv.org/abs/2003.07369}{{\ttfamily
  2003.07369}}].

\bibitem{DeRoeck:2020ntj}
A.~De~Roeck, D.~Kim, Z.G.~Moghaddam, J.-C.~Park, S.~Shin and L.H.~Whitehead,
  \emph{{Probing Energetic Light Dark Matter with Multi-Particle Tracks
  Signatures at DUNE}},
  \href{https://doi.org/10.1007/JHEP11(2020)043}{\emph{JHEP} {\bfseries 11}
  (2020) 043} [\href{https://arxiv.org/abs/2005.08979}{{\ttfamily
  2005.08979}}].

\bibitem{Alhazmi:2020fju}
H.~Alhazmi, D.~Kim, K.~Kong, G.~Mohlabeng, J.-C.~Park and S.~Shin,
  \emph{{Implications of the XENON1T Excess on the Dark Matter
  Interpretation}}, \href{https://doi.org/10.1007/JHEP05(2021)055}{\emph{JHEP}
  {\bfseries 05} (2021) 055}
  [\href{https://arxiv.org/abs/2006.16252}{{\ttfamily 2006.16252}}].

\bibitem{Bringmann:2018cvk}
T.~Bringmann and M.~Pospelov, \emph{{Novel direct detection constraints on
  light dark matter}},
  \href{https://doi.org/10.1103/PhysRevLett.122.171801}{\emph{Phys. Rev. Lett.}
  {\bfseries 122} (2019) 171801}
  [\href{https://arxiv.org/abs/1810.10543}{{\ttfamily 1810.10543}}].

\bibitem{Ema:2018bih}
Y.~Ema, F.~Sala and R.~Sato, \emph{{Light Dark Matter at Neutrino
  Experiments}},
  \href{https://doi.org/10.1103/PhysRevLett.122.181802}{\emph{Phys. Rev. Lett.}
  {\bfseries 122} (2019) 181802}
  [\href{https://arxiv.org/abs/1811.00520}{{\ttfamily 1811.00520}}].

\bibitem{Cappiello:2019qsw}
C.V.~Cappiello and J.F.~Beacom, \emph{{Strong New Limits on Light Dark Matter
  from Neutrino Experiments}},
  \href{https://doi.org/10.1103/PhysRevD.104.069901}{\emph{Phys. Rev. D}
  {\bfseries 100} (2019) 103011}
  [\href{https://arxiv.org/abs/1906.11283}{{\ttfamily 1906.11283}}].

\bibitem{Dent:2019krz}
J.B.~Dent, B.~Dutta, J.L.~Newstead and I.M.~Shoemaker, \emph{{Bounds on Cosmic
  Ray-Boosted Dark Matter in Simplified Models and its Corresponding
  Neutrino-Floor}},
  \href{https://doi.org/10.1103/PhysRevD.101.116007}{\emph{Phys. Rev. D}
  {\bfseries 101} (2020) 116007}
  [\href{https://arxiv.org/abs/1907.03782}{{\ttfamily 1907.03782}}].

\bibitem{Wang:2019jtk}
W.~Wang, L.~Wu, J.M.~Yang, H.~Zhou and B.~Zhu, \emph{{Cosmic ray boosted
  sub-GeV gravitationally interacting dark matter in direct detection}},
  \href{https://doi.org/10.1007/JHEP12(2020)072}{\emph{JHEP} {\bfseries 12}
  (2020) 072} [\href{https://arxiv.org/abs/1912.09904}{{\ttfamily
  1912.09904}}].

\bibitem{Cao:2020bwd}
Q.-H.~Cao, R.~Ding and Q.-F.~Xiang, \emph{{Searching for sub-MeV boosted dark
  matter from xenon electron direct detection}},
  \href{https://doi.org/10.1088/1674-1137/abe195}{\emph{Chin. Phys. C}
  {\bfseries 45} (2021) 045002}
  [\href{https://arxiv.org/abs/2006.12767}{{\ttfamily 2006.12767}}].

\bibitem{Jho:2020sku}
Y.~Jho, J.-C.~Park, S.C.~Park and P.-Y.~Tseng, \emph{{Leptonic New Force and
  Cosmic-ray Boosted Dark Matter for the XENON1T Excess}},
  \href{https://doi.org/10.1016/j.physletb.2020.135863}{\emph{Phys. Lett. B}
  {\bfseries 811} (2020) 135863}
  [\href{https://arxiv.org/abs/2006.13910}{{\ttfamily 2006.13910}}].

\bibitem{Cho:2020mnc}
W.~Cho, K.-Y.~Choi and S.M.~Yoo, \emph{{Searching for boosted dark matter
  mediated by a new gauge boson}},
  \href{https://doi.org/10.1103/PhysRevD.102.095010}{\emph{Phys. Rev. D}
  {\bfseries 102} (2020) 095010}
  [\href{https://arxiv.org/abs/2007.04555}{{\ttfamily 2007.04555}}].

\bibitem{Dent:2020syp}
J.B.~Dent, B.~Dutta, J.L.~Newstead, I.M.~Shoemaker and N.T.~Arellano,
  \emph{{Present and future status of light dark matter models from cosmic-ray
  electron upscattering}},
  \href{https://doi.org/10.1103/PhysRevD.103.095015}{\emph{Phys. Rev. D}
  {\bfseries 103} (2021) 095015}
  [\href{https://arxiv.org/abs/2010.09749}{{\ttfamily 2010.09749}}].

\bibitem{Xia:2021vbz}
C.~Xia, Y.-H.~Xu and Y.-F.~Zhou, \emph{{Production and attenuation of
  cosmic-ray boosted dark matter}},
  \href{https://doi.org/10.1088/1475-7516/2022/02/028}{\emph{JCAP} {\bfseries
  02} (2022) 028} [\href{https://arxiv.org/abs/2111.05559}{{\ttfamily
  2111.05559}}].

\bibitem{Ghosh:2021vkt}
D.~Ghosh, A.~Guha and D.~Sachdeva, \emph{{Exclusion limits on dark
  matter-neutrino scattering cross section}},
  \href{https://doi.org/10.1103/PhysRevD.105.103029}{\emph{Phys. Rev. D}
  {\bfseries 105} (2022) 103029}
  [\href{https://arxiv.org/abs/2110.00025}{{\ttfamily 2110.00025}}].

\bibitem{Bardhan:2022bdg}
D.~Bardhan, S.~Bhowmick, D.~Ghosh, A.~Guha and D.~Sachdeva, \emph{{Bounds on
  boosted dark matter from direct detection: The role of energy-dependent cross
  sections}}, \href{https://doi.org/10.1103/PhysRevD.107.015010}{\emph{Phys.
  Rev. D} {\bfseries 107} (2023) 015010}
  [\href{https://arxiv.org/abs/2208.09405}{{\ttfamily 2208.09405}}].

\bibitem{Alvey:2022pad}
J.~Alvey, T.~Bringmann and H.~Kolesova, \emph{{No room to hide: implications of
  cosmic-ray upscattering for GeV-scale dark matter}},
  \href{https://doi.org/10.1007/JHEP01(2023)123}{\emph{JHEP} {\bfseries 01}
  (2023) 123} [\href{https://arxiv.org/abs/2209.03360}{{\ttfamily
  2209.03360}}].

\bibitem{Maity:2022exk}
T.N.~Maity and R.~Laha, \emph{{Cosmic-ray boosted dark matter in Xe-based
  direct detection experiments}},
  \href{https://doi.org/10.1140/epjc/s10052-024-12464-8}{\emph{Eur. Phys. J. C}
  {\bfseries 84} (2024) 117}
  [\href{https://arxiv.org/abs/2210.01815}{{\ttfamily 2210.01815}}].

\bibitem{Bell:2023sdq}
N.F.~Bell, J.L.~Newstead and I.~Shaukat-Ali, \emph{{Cosmic-ray dark matter
  confronted by constraints on new light mediators}},
  \href{https://arxiv.org/abs/2309.11003}{{\ttfamily 2309.11003}}.

\bibitem{Jho:2021rmn}
Y.~Jho, J.-C.~Park, S.C.~Park and P.-Y.~Tseng, \emph{{Cosmic-Neutrino-Boosted
  Dark Matter ($\nu$BDM)}},  \href{https://arxiv.org/abs/2101.11262}{{\ttfamily
  2101.11262}}.

\bibitem{Das:2021lcr}
A.~Das and M.~Sen, \emph{{Boosted dark matter from diffuse supernova
  neutrinos}}, \href{https://doi.org/10.1103/PhysRevD.104.075029}{\emph{Phys.
  Rev. D} {\bfseries 104} (2021) 075029}
  [\href{https://arxiv.org/abs/2104.00027}{{\ttfamily 2104.00027}}].

\bibitem{Chao:2021orr}
W.~Chao, T.~Li and J.~Liao, \emph{{Connecting Primordial Black Hole to boosted
  sub-GeV Dark Matter through neutrino}},
  \href{https://arxiv.org/abs/2108.05608}{{\ttfamily 2108.05608}}.

\bibitem{Lin:2022dbl}
Y.-H.~Lin, W.-H.~Wu, M.-R.~Wu and H.T.-K.~Wong, \emph{{Searching for Afterglow:
  Light Dark Matter Boosted by Supernova Neutrinos}},
  \href{https://doi.org/10.1103/PhysRevLett.130.111002}{\emph{Phys. Rev. Lett.}
  {\bfseries 130} (2023) 111002}
  [\href{https://arxiv.org/abs/2206.06864}{{\ttfamily 2206.06864}}].

\bibitem{DeRomeri:2023ytt}
V.~De~Romeri, A.~Majumdar, D.K.~Papoulias and R.~Srivastava, \emph{{XENONnT and
  LUX-ZEPLIN constraints on DSNB-boosted dark matter}},
  \href{https://arxiv.org/abs/2309.04117}{{\ttfamily 2309.04117}}.

\bibitem{Super-Kamiokande:2017dch}
{\scshape Super-Kamiokande} collaboration, \emph{{Search for Boosted Dark
  Matter Interacting With Electrons in Super-Kamiokande}},
  \href{https://doi.org/10.1103/PhysRevLett.120.221301}{\emph{Phys. Rev. Lett.}
  {\bfseries 120} (2018) 221301}
  [\href{https://arxiv.org/abs/1711.05278}{{\ttfamily 1711.05278}}].

\bibitem{COSINE-100:2018ged}
{\scshape COSINE-100} collaboration, \emph{{First Direct Search for Inelastic
  Boosted Dark Matter with COSINE-100}},
  \href{https://doi.org/10.1103/PhysRevLett.122.131802}{\emph{Phys. Rev. Lett.}
  {\bfseries 122} (2019) 131802}
  [\href{https://arxiv.org/abs/1811.09344}{{\ttfamily 1811.09344}}].

\bibitem{PandaX-II:2021kai}
{\scshape PandaX-II} collaboration, \emph{{Search for Cosmic-Ray Boosted
  Sub-GeV Dark Matter at the PandaX-II Experiment}},
  \href{https://doi.org/10.1103/PhysRevLett.128.171801}{\emph{Phys. Rev. Lett.}
  {\bfseries 128} (2022) 171801}
  [\href{https://arxiv.org/abs/2112.08957}{{\ttfamily 2112.08957}}].

\bibitem{CDEX:2022fig}
{\scshape CDEX} collaboration, \emph{{Constraints on sub-GeV Dark Matter
  Boosted by Cosmic Rays from CDEX-10 Experiment at the China Jinping
  Underground Laboratory}},  \href{https://arxiv.org/abs/2201.01704}{{\ttfamily
  2201.01704}}.

\bibitem{Super-Kamiokande:2022ncz}
{\scshape Super-Kamiokande} collaboration, \emph{{Search for Cosmic-Ray Boosted
  Sub-GeV Dark Matter Using Recoil Protons at Super-Kamiokande}},
  \href{https://doi.org/10.1103/PhysRevLett.130.031802}{\emph{Phys. Rev. Lett.}
  {\bfseries 130} (2023) 031802}
  [\href{https://arxiv.org/abs/2209.14968}{{\ttfamily 2209.14968}}].

\bibitem{Workman:2022ynf}
{\scshape Particle Data Group} collaboration, \emph{{Review of Particle
  Physics}}, \href{https://doi.org/10.1093/ptep/ptac097}{\emph{PTEP} {\bfseries
  2022} (2022) 083C01}.

\bibitem{PhysRevD.98.030001}
{\scshape Particle Data Group} collaboration, \emph{Review of particle
  physics}, \href{https://doi.org/10.1103/PhysRevD.98.030001}{\emph{Phys. Rev.
  D} {\bfseries 98} (2018) 030001}.

\bibitem{Huh:2007zw}
J.-H.~Huh, J.E.~Kim, J.-C.~Park and S.C.~Park, \emph{{Galactic 511 keV line
  from MeV milli-charged dark matter}},
  \href{https://doi.org/10.1103/PhysRevD.77.123503}{\emph{Phys. Rev. D}
  {\bfseries 77} (2008) 123503}
  [\href{https://arxiv.org/abs/0711.3528}{{\ttfamily 0711.3528}}].

\bibitem{Pospelov:2007mp}
M.~Pospelov, A.~Ritz and M.B.~Voloshin, \emph{{Secluded WIMP Dark Matter}},
  \href{https://doi.org/10.1016/j.physletb.2008.02.052}{\emph{Phys. Lett. B}
  {\bfseries 662} (2008) 53} [\href{https://arxiv.org/abs/0711.4866}{{\ttfamily
  0711.4866}}].

\bibitem{Chun:2010ve}
E.J.~Chun, J.-C.~Park and S.~Scopel, \emph{{Dark matter and a new gauge boson
  through kinetic mixing}},
  \href{https://doi.org/10.1007/JHEP02(2011)100}{\emph{JHEP} {\bfseries 02}
  (2011) 100} [\href{https://arxiv.org/abs/1011.3300}{{\ttfamily 1011.3300}}].

\bibitem{Park:2023hsp}
J.-C.~Park and G.~Tomar, \emph{{Probing non-standard neutrino interactions with
  interference: insights from dark matter and neutrino experiments}},
  \href{https://doi.org/10.1088/1475-7516/2023/08/025}{\emph{JCAP} {\bfseries
  08} (2023) 025} [\href{https://arxiv.org/abs/2305.10836}{{\ttfamily
  2305.10836}}].

\bibitem{Longair:1992ze}
M.S.~Longair, ed., \emph{{High-energy astrophysics. Vol. 1: Particles, photons
  and their detection}}, "Cambridge University Press" (2011).

\bibitem{Boschini:2018zdv}
M.J.~Boschini et~al., \emph{{HelMod in the works: from direct observations to
  the local interstellar spectrum of cosmic-ray electrons}},
  \href{https://doi.org/10.3847/1538-4357/aaa75e}{\emph{Astrophys. J.}
  {\bfseries 854} (2018) 94}
  [\href{https://arxiv.org/abs/1801.04059}{{\ttfamily 1801.04059}}].

\bibitem{Fermi-LAT:2011baq}
{\scshape Fermi-LAT} collaboration, \emph{{Measurement of separate cosmic-ray
  electron and positron spectra with the Fermi Large Area Telescope}},
  \href{https://doi.org/10.1103/PhysRevLett.108.011103}{\emph{Phys. Rev. Lett.}
  {\bfseries 108} (2012) 011103}
  [\href{https://arxiv.org/abs/1109.0521}{{\ttfamily 1109.0521}}].

\bibitem{Fermi-LAT:2009yfs}
{\scshape Fermi-LAT} collaboration, \emph{{Measurement of the Cosmic Ray e+
  plus e- spectrum from 20 GeV to 1 TeV with the Fermi Large Area Telescope}},
  \href{https://doi.org/10.1103/PhysRevLett.102.181101}{\emph{Phys. Rev. Lett.}
  {\bfseries 102} (2009) 181101}
  [\href{https://arxiv.org/abs/0905.0025}{{\ttfamily 0905.0025}}].

\bibitem{Fermi-LAT:2010fit}
{\scshape Fermi-LAT} collaboration, \emph{{Fermi LAT observations of cosmic-ray
  electrons from 7 GeV to 1 TeV}},
  \href{https://doi.org/10.1103/PhysRevD.82.092004}{\emph{Phys. Rev. D}
  {\bfseries 82} (2010) 092004}
  [\href{https://arxiv.org/abs/1008.3999}{{\ttfamily 1008.3999}}].

\bibitem{Fermi-LAT:2017bpc}
{\scshape Fermi-LAT} collaboration, \emph{{Cosmic-ray electron-positron
  spectrum from 7 GeV to 2 TeV with the Fermi Large Area Telescope}},
  \href{https://doi.org/10.1103/PhysRevD.95.082007}{\emph{Phys. Rev. D}
  {\bfseries 95} (2017) 082007}
  [\href{https://arxiv.org/abs/1704.07195}{{\ttfamily 1704.07195}}].

\bibitem{AMS:2014gdf}
{\scshape AMS} collaboration, \emph{{Precision Measurement of the ($e^+ + e^-$)
  Flux in Primary Cosmic Rays from 0.5 GeV to 1 TeV with the Alpha Magnetic
  Spectrometer on the International Space Station}},
  \href{https://doi.org/10.1103/PhysRevLett.113.221102}{\emph{Phys. Rev. Lett.}
  {\bfseries 113} (2014) 221102}.

\bibitem{PAMELA:2011bbe}
{\scshape PAMELA} collaboration, \emph{{The cosmic-ray electron flux measured
  by the PAMELA experiment between 1 and 625 GeV}},
  \href{https://doi.org/10.1103/PhysRevLett.106.201101}{\emph{Phys. Rev. Lett.}
  {\bfseries 106} (2011) 201101}
  [\href{https://arxiv.org/abs/1103.2880}{{\ttfamily 1103.2880}}].

\bibitem{CALET:2017uxd}
{\scshape CALET} collaboration, \emph{{Energy Spectrum of Cosmic-Ray Electron
  and Positron from 10 GeV to 3 TeV Observed with the Calorimetric Electron
  Telescope on the International Space Station}},
  \href{https://doi.org/10.1103/PhysRevLett.119.181101}{\emph{Phys. Rev. Lett.}
  {\bfseries 119} (2017) 181101}
  [\href{https://arxiv.org/abs/1712.01711}{{\ttfamily 1712.01711}}].

\bibitem{Cummings:2016pdr}
A.C.~Cummings, E.C.~Stone, B.C.~Heikkila, N.~Lal, W.R.~Webber, G.~J\'ohannesson
  et~al., \emph{{Galactic Cosmic Rays in the Local Interstellar Medium: Voyager
  1 Observations and Model Results}},
  \href{https://doi.org/10.3847/0004-637X/831/1/18}{\emph{Astrophys. J.}
  {\bfseries 831} (2016) 18}.

\bibitem{Stone:2013}
E.C.~Stone, A.C.~Cummings, F.B.~McDonald, B.C.~Heikkila, N.~Lal and
  W.R.~Webber, \emph{Voyager 1 observes low-energy galactic cosmic rays in a
  region depleted of heliospheric ions},
  \href{https://doi.org/10.1126/science.1236408}{\emph{Science} {\bfseries 341}
  (2013) 150}
  [\href{https://arxiv.org/abs/https://www.science.org/doi/pdf/10.1126/science.1236408}{{\ttfamily
  https://www.science.org/doi/pdf/10.1126/science.1236408}}].

\bibitem{Boschini:2017fxq}
M.J.~Boschini et~al., \emph{{Solution of heliospheric propagation: unveiling
  the local interstellar spectra of cosmic ray species}},
  \href{https://doi.org/10.3847/1538-4357/aa6e4f}{\emph{Astrophys. J.}
  {\bfseries 840} (2017) 115}
  [\href{https://arxiv.org/abs/1704.06337}{{\ttfamily 1704.06337}}].

\bibitem{DellaTorre:2016jjf}
S.~Della~Torre et~al., \emph{{From Observations near the Earth to the Local
  Interstellar Spectra}},  in \emph{{25th European Cosmic Ray Symposium}}, 12,
  2016 [\href{https://arxiv.org/abs/1701.02363}{{\ttfamily 1701.02363}}].

\bibitem{Fabbrichesi:2020wbt}
M.~Fabbrichesi, E.~Gabrielli and G.~Lanfranchi, \emph{{The Dark Photon}},
  \href{https://arxiv.org/abs/2005.01515}{{\ttfamily 2005.01515}}.

\bibitem{Gu:2017gle}
P.-H.~Gu and X.-G.~He, \emph{{Electrophilic dark matter with dark photon: from
  DAMPE to direct detection}},
  \href{https://doi.org/10.1016/j.physletb.2018.01.057}{\emph{Phys. Lett. B}
  {\bfseries 778} (2018) 292}
  [\href{https://arxiv.org/abs/1711.11000}{{\ttfamily 1711.11000}}].

\bibitem{Bauer:2018onh}
M.~Bauer, P.~Foldenauer and J.~Jaeckel, \emph{{Hunting All the Hidden
  Photons}}, \href{https://doi.org/10.1007/JHEP07(2018)094}{\emph{JHEP}
  {\bfseries 07} (2018) 094}
  [\href{https://arxiv.org/abs/1803.05466}{{\ttfamily 1803.05466}}].

\bibitem{Inan:2021dir}
S.C.~\.Inan and A.V.~Kisselev, \emph{{Search for invisible dark photon in
  $\gamma e$ scattering at future lepton colliders}},
  \href{https://doi.org/10.1140/epjc/s10052-022-10552-1}{\emph{Eur. Phys. J. C}
  {\bfseries 82} (2022) 592}
  [\href{https://arxiv.org/abs/2112.13070}{{\ttfamily 2112.13070}}].

\bibitem{Filippi:2020kii}
A.~Filippi and M.~De~Napoli, \emph{{Searching in the dark: the hunt for the
  dark photon}}, \href{https://doi.org/10.1016/j.revip.2020.100042}{\emph{Rev.
  Phys.} {\bfseries 5} (2020) 100042}
  [\href{https://arxiv.org/abs/2006.04640}{{\ttfamily 2006.04640}}].

\bibitem{Gninenko:2020hbd}
S.N.~Gninenko, N.V.~Krasnikov and V.A.~Matveev, \emph{{Search for dark sector
  physics with NA64}},
  \href{https://doi.org/10.1134/S1063779620050044}{\emph{Phys. Part. Nucl.}
  {\bfseries 51} (2020) 829}
  [\href{https://arxiv.org/abs/2003.07257}{{\ttfamily 2003.07257}}].

\bibitem{Mohlabeng:2019vrz}
G.~Mohlabeng, \emph{{Revisiting the dark photon explanation of the muon
  anomalous magnetic moment}},
  \href{https://doi.org/10.1103/PhysRevD.99.115001}{\emph{Phys. Rev. D}
  {\bfseries 99} (2019) 115001}
  [\href{https://arxiv.org/abs/1902.05075}{{\ttfamily 1902.05075}}].

\bibitem{Andreev:2021fzd}
Y.M.~Andreev et~al., \emph{{Improved exclusion limit for light dark matter from
  e+e- annihilation in NA64}},
  \href{https://doi.org/10.1103/PhysRevD.104.L091701}{\emph{Phys. Rev. D}
  {\bfseries 104} (2021) L091701}
  [\href{https://arxiv.org/abs/2108.04195}{{\ttfamily 2108.04195}}].

\bibitem{Banerjee:2019pds}
D.~Banerjee et~al., \emph{{Dark matter search in missing energy events with
  NA64}}, \href{https://doi.org/10.1103/PhysRevLett.123.121801}{\emph{Phys.
  Rev. Lett.} {\bfseries 123} (2019) 121801}
  [\href{https://arxiv.org/abs/1906.00176}{{\ttfamily 1906.00176}}].

\bibitem{NA62:2019meo}
{\scshape NA62} collaboration, \emph{{Search for production of an invisible
  dark photon in $\pi^0$ decays}},
  \href{https://doi.org/10.1007/JHEP05(2019)182}{\emph{JHEP} {\bfseries 05}
  (2019) 182} [\href{https://arxiv.org/abs/1903.08767}{{\ttfamily
  1903.08767}}].

\bibitem{Filippi:2019lfq}
{\scshape BABAR} collaboration, \emph{{Dark Photon Studies at BABAR}},
  \href{https://doi.org/10.1051/epjconf/201921806001}{\emph{EPJ Web Conf.}
  {\bfseries 218} (2019) 06001}.

\bibitem{BaBar:2017tiz}
{\scshape BaBar} collaboration, \emph{{Search for Invisible Decays of a Dark
  Photon Produced in ${e}^{+}{e}^{-}$ Collisions at BaBar}},
  \href{https://doi.org/10.1103/PhysRevLett.119.131804}{\emph{Phys. Rev. Lett.}
  {\bfseries 119} (2017) 131804}
  [\href{https://arxiv.org/abs/1702.03327}{{\ttfamily 1702.03327}}].

\bibitem{Essig:2013vha}
R.~Essig, J.~Mardon, M.~Papucci, T.~Volansky and Y.-M.~Zhong,
  \emph{{Constraining Light Dark Matter with Low-Energy $e^+e^-$ Colliders}},
  \href{https://doi.org/10.1007/JHEP11(2013)167}{\emph{JHEP} {\bfseries 11}
  (2013) 167} [\href{https://arxiv.org/abs/1309.5084}{{\ttfamily 1309.5084}}].

\bibitem{NA64:2019auh}
{\scshape NA64} collaboration, \emph{{Improved limits on a hypothetical X(16.7)
  boson and a dark photon decaying into $e^+e^-$ pairs}},
  \href{https://doi.org/10.1103/PhysRevD.101.071101}{\emph{Phys. Rev. D}
  {\bfseries 101} (2020) 071101}
  [\href{https://arxiv.org/abs/1912.11389}{{\ttfamily 1912.11389}}].

\bibitem{NA482:2015wmo}
{\scshape NA48/2} collaboration, \emph{{Search for the dark photon in $\pi^0$
  decays}}, \href{https://doi.org/10.1016/j.physletb.2015.04.068}{\emph{Phys.
  Lett. B} {\bfseries 746} (2015) 178}
  [\href{https://arxiv.org/abs/1504.00607}{{\ttfamily 1504.00607}}].

\bibitem{WASA-at-COSY:2013zom}
{\scshape WASA-at-COSY} collaboration, \emph{{Search for a dark photon in the
  $\pi^0 \to e^+e^-\gamma$ decay}},
  \href{https://doi.org/10.1016/j.physletb.2013.08.055}{\emph{Phys. Lett. B}
  {\bfseries 726} (2013) 187}
  [\href{https://arxiv.org/abs/1304.0671}{{\ttfamily 1304.0671}}].

\bibitem{BESIII:2017fwv}
{\scshape BESIII} collaboration, \emph{{Dark Photon Search in the Mass Range
  Between 1.5 and 3.4 GeV/$c^2$}},
  \href{https://doi.org/10.1016/j.physletb.2017.09.067}{\emph{Phys. Lett. B}
  {\bfseries 774} (2017) 252}
  [\href{https://arxiv.org/abs/1705.04265}{{\ttfamily 1705.04265}}].

\bibitem{BaBar:2014zli}
{\scshape BaBar} collaboration, \emph{{Search for a Dark Photon in $e^+e^-$
  Collisions at BaBar}},
  \href{https://doi.org/10.1103/PhysRevLett.113.201801}{\emph{Phys. Rev. Lett.}
  {\bfseries 113} (2014) 201801}
  [\href{https://arxiv.org/abs/1406.2980}{{\ttfamily 1406.2980}}].

\bibitem{LSND:1997vqj}
{\scshape LSND} collaboration, \emph{{Evidence for muon-neutrino
  ---\ensuremath{>} electron-neutrino oscillations from pion decay in flight
  neutrinos}}, \href{https://doi.org/10.1103/PhysRevC.58.2489}{\emph{Phys. Rev.
  C} {\bfseries 58} (1998) 2489}
  [\href{https://arxiv.org/abs/nucl-ex/9706006}{{\ttfamily nucl-ex/9706006}}].

\bibitem{HADES:2013nab}
{\scshape HADES} collaboration, \emph{{Searching a Dark Photon with HADES}},
  \href{https://doi.org/10.1016/j.physletb.2014.02.035}{\emph{Phys. Lett. B}
  {\bfseries 731} (2014) 265}
  [\href{https://arxiv.org/abs/1311.0216}{{\ttfamily 1311.0216}}].

\bibitem{KLOE-2:2012lii}
{\scshape KLOE-2} collaboration, \emph{{Limit on the production of a light
  vector gauge boson in phi meson decays with the KLOE detector}},
  \href{https://doi.org/10.1016/j.physletb.2013.01.067}{\emph{Phys. Lett. B}
  {\bfseries 720} (2013) 111}
  [\href{https://arxiv.org/abs/1210.3927}{{\ttfamily 1210.3927}}].

\bibitem{KLOE-2:2014qxg}
{\scshape KLOE-2} collaboration, \emph{{Search for light vector boson
  production in $e^+e^- \rightarrow \mu^+ \mu^- \gamma$ interactions with the
  KLOE experiment}},
  \href{https://doi.org/10.1016/j.physletb.2014.08.005}{\emph{Phys. Lett. B}
  {\bfseries 736} (2014) 459}
  [\href{https://arxiv.org/abs/1404.7772}{{\ttfamily 1404.7772}}].

\bibitem{Anastasi:2015qla}
A.~Anastasi et~al., \emph{{Limit on the production of a low-mass vector boson
  in $\mathrm{e}^{+}\mathrm{e}^{-} \to \mathrm{U}\gamma$, $\mathrm{U} \to
  \mathrm{e}^{+}\mathrm{e}^{-}$ with the KLOE experiment}},
  \href{https://doi.org/10.1016/j.physletb.2015.10.003}{\emph{Phys. Lett. B}
  {\bfseries 750} (2015) 633}
  [\href{https://arxiv.org/abs/1509.00740}{{\ttfamily 1509.00740}}].

\bibitem{Riordan:1987aw}
E.M.~Riordan et~al., \emph{{A Search for Short Lived Axions in an Electron Beam
  Dump Experiment}},
  \href{https://doi.org/10.1103/PhysRevLett.59.755}{\emph{Phys. Rev. Lett.}
  {\bfseries 59} (1987) 755}.

\bibitem{Bjorken:1988as}
J.D.~Bjorken, S.~Ecklund, W.R.~Nelson, A.~Abashian, C.~Church, B.~Lu et~al.,
  \emph{{Search for Neutral Metastable Penetrating Particles Produced in the
  SLAC Beam Dump}}, \href{https://doi.org/10.1103/PhysRevD.38.3375}{\emph{Phys.
  Rev. D} {\bfseries 38} (1988) 3375}.

\bibitem{Batell:2014mga}
B.~Batell, R.~Essig and Z.~Surujon, \emph{{Strong Constraints on Sub-GeV Dark
  Sectors from SLAC Beam Dump E137}},
  \href{https://doi.org/10.1103/PhysRevLett.113.171802}{\emph{Phys. Rev. Lett.}
  {\bfseries 113} (2014) 171802}
  [\href{https://arxiv.org/abs/1406.2698}{{\ttfamily 1406.2698}}].

\bibitem{Marsicano:2018krp}
L.~Marsicano, M.~Battaglieri, M.~Bondi', C.D.R.~Carvajal, A.~Celentano,
  M.~De~Napoli et~al., \emph{{Dark photon production through positron
  annihilation in beam-dump experiments}},
  \href{https://doi.org/10.1103/PhysRevD.98.015031}{\emph{Phys. Rev. D}
  {\bfseries 98} (2018) 015031}
  [\href{https://arxiv.org/abs/1802.03794}{{\ttfamily 1802.03794}}].

\bibitem{Bross:1989mp}
A.~Bross, M.~Crisler, S.H.~Pordes, J.~Volk, S.~Errede and J.~Wrbanek, \emph{{A
  Search for Shortlived Particles Produced in an Electron Beam Dump}},
  \href{https://doi.org/10.1103/PhysRevLett.67.2942}{\emph{Phys. Rev. Lett.}
  {\bfseries 67} (1991) 2942}.

\bibitem{Blumlein:2011mv}
J.~Blumlein and J.~Brunner, \emph{{New Exclusion Limits for Dark Gauge Forces
  from Beam-Dump Data}},
  \href{https://doi.org/10.1016/j.physletb.2011.05.046}{\emph{Phys. Lett. B}
  {\bfseries 701} (2011) 155}
  [\href{https://arxiv.org/abs/1104.2747}{{\ttfamily 1104.2747}}].

\bibitem{Blumlein:2013cua}
J.~Bl\"umlein and J.~Brunner, \emph{{New Exclusion Limits on Dark Gauge Forces
  from Proton Bremsstrahlung in Beam-Dump Data}},
  \href{https://doi.org/10.1016/j.physletb.2014.02.029}{\emph{Phys. Lett. B}
  {\bfseries 731} (2014) 320}
  [\href{https://arxiv.org/abs/1311.3870}{{\ttfamily 1311.3870}}].

\bibitem{Gninenko:2012eq}
S.N.~Gninenko, \emph{{Constraints on sub-GeV hidden sector gauge bosons from a
  search for heavy neutrino decays}},
  \href{https://doi.org/10.1016/j.physletb.2012.06.002}{\emph{Phys. Lett. B}
  {\bfseries 713} (2012) 244}
  [\href{https://arxiv.org/abs/1204.3583}{{\ttfamily 1204.3583}}].

\bibitem{APEX:2011dww}
{\scshape APEX} collaboration, \emph{{Search for a New Gauge Boson in
  Electron-Nucleus Fixed-Target Scattering by the APEX Experiment}},
  \href{https://doi.org/10.1103/PhysRevLett.107.191804}{\emph{Phys. Rev. Lett.}
  {\bfseries 107} (2011) 191804}
  [\href{https://arxiv.org/abs/1108.2750}{{\ttfamily 1108.2750}}].

\bibitem{Merkel:2014avp}
H.~Merkel et~al., \emph{{Search at the Mainz Microtron for Light Massive Gauge
  Bosons Relevant for the Muon g-2 Anomaly}},
  \href{https://doi.org/10.1103/PhysRevLett.112.221802}{\emph{Phys. Rev. Lett.}
  {\bfseries 112} (2014) 221802}
  [\href{https://arxiv.org/abs/1404.5502}{{\ttfamily 1404.5502}}].

\bibitem{Perdrisat:2006hj}
C.F.~Perdrisat, V.~Punjabi and M.~Vanderhaeghen, \emph{{Nucleon Electromagnetic
  Form Factors}}, \href{https://doi.org/10.1016/j.ppnp.2007.05.001}{\emph{Prog.
  Part. Nucl. Phys.} {\bfseries 59} (2007) 694}
  [\href{https://arxiv.org/abs/hep-ph/0612014}{{\ttfamily hep-ph/0612014}}].

\bibitem{Lei:2020mii}
Z.-H.~Lei, J.~Tang and B.-L.~Zhang, \emph{{Constraints on cosmic-ray boosted
  dark matter in CDEX-10 *}},
  \href{https://doi.org/10.1088/1674-1137/ac68da}{\emph{Chin. Phys. C}
  {\bfseries 46} (2022) 085103}
  [\href{https://arxiv.org/abs/2008.07116}{{\ttfamily 2008.07116}}].

\bibitem{XENON:2020rca}
{\scshape XENON} collaboration, \emph{{Excess electronic recoil events in
  XENON1T}}, \href{https://doi.org/10.1103/PhysRevD.102.072004}{\emph{Phys.
  Rev. D} {\bfseries 102} (2020) 072004}
  [\href{https://arxiv.org/abs/2006.09721}{{\ttfamily 2006.09721}}].

\bibitem{Essig:2011nj}
R.~Essig, J.~Mardon and T.~Volansky, \emph{{Direct Detection of Sub-GeV Dark
  Matter}}, \href{https://doi.org/10.1103/PhysRevD.85.076007}{\emph{Phys. Rev.
  D} {\bfseries 85} (2012) 076007}
  [\href{https://arxiv.org/abs/1108.5383}{{\ttfamily 1108.5383}}].

\bibitem{PandaX-II:2021nsg}
{\scshape PandaX-II} collaboration, \emph{{Search for Light Dark
  Matter-Electron Scatterings in the PandaX-II Experiment}},
  \href{https://doi.org/10.1103/PhysRevLett.126.211803}{\emph{Phys. Rev. Lett.}
  {\bfseries 126} (2021) 211803}
  [\href{https://arxiv.org/abs/2101.07479}{{\ttfamily 2101.07479}}].

\bibitem{Essig:2017kqs}
R.~Essig, T.~Volansky and T.-T.~Yu, \emph{{New Constraints and Prospects for
  sub-GeV Dark Matter Scattering off Electrons in Xenon}},
  \href{https://doi.org/10.1103/PhysRevD.96.043017}{\emph{Phys. Rev. D}
  {\bfseries 96} (2017) 043017}
  [\href{https://arxiv.org/abs/1703.00910}{{\ttfamily 1703.00910}}].

\bibitem{XENON:2019gfn}
{\scshape XENON} collaboration, \emph{{Light Dark Matter Search with Ionization
  Signals in XENON1T}},
  \href{https://doi.org/10.1103/PhysRevLett.123.251801}{\emph{Phys. Rev. Lett.}
  {\bfseries 123} (2019) 251801}
  [\href{https://arxiv.org/abs/1907.11485}{{\ttfamily 1907.11485}}].

\bibitem{DAMIC:2019dcn}
{\scshape DAMIC} collaboration, \emph{{Constraints on Light Dark Matter
  Particles Interacting with Electrons from DAMIC at SNOLAB}},
  \href{https://doi.org/10.1103/PhysRevLett.123.181802}{\emph{Phys. Rev. Lett.}
  {\bfseries 123} (2019) 181802}
  [\href{https://arxiv.org/abs/1907.12628}{{\ttfamily 1907.12628}}].

\bibitem{Griffin:2019mvc}
S.M.~Griffin, K.~Inzani, T.~Trickle, Z.~Zhang and K.M.~Zurek,
  \emph{{Multichannel direct detection of light dark matter: Target
  comparison}}, \href{https://doi.org/10.1103/PhysRevD.101.055004}{\emph{Phys.
  Rev. D} {\bfseries 101} (2020) 055004}
  [\href{https://arxiv.org/abs/1910.10716}{{\ttfamily 1910.10716}}].

\bibitem{Taufertshofer:2023rgq}
N.~Taufertsh\"ofer, M.~Garcia-Sciveres and S.M.~Griffin, \emph{{Broad-Range
  Directional Detection of Light Dark Matter in Cryogenic Ice}},
  \href{https://arxiv.org/abs/2301.04778}{{\ttfamily 2301.04778}}.

\bibitem{Bellomo:2022qbx}
N.~Bellomo, K.V.~Berghaus and K.K.~Boddy, \emph{{Dark matter freeze-in produces
  large post-inflationary isocurvature}},
  \href{https://arxiv.org/abs/2210.15691}{{\ttfamily 2210.15691}}.

\bibitem{Emken:2018run}
T.~Emken and C.~Kouvaris, \emph{{How blind are underground and surface
  detectors to strongly interacting Dark Matter?}},
  \href{https://doi.org/10.1103/PhysRevD.97.115047}{\emph{Phys. Rev. D}
  {\bfseries 97} (2018) 115047}
  [\href{https://arxiv.org/abs/1802.04764}{{\ttfamily 1802.04764}}].

\bibitem{Emken:2017qmp}
T.~Emken and C.~Kouvaris, \emph{{DaMaSCUS: The Impact of Underground
  Scatterings on Direct Detection of Light Dark Matter}},
  \href{https://doi.org/10.1088/1475-7516/2017/10/031}{\emph{JCAP} {\bfseries
  10} (2017) 031} [\href{https://arxiv.org/abs/1706.02249}{{\ttfamily
  1706.02249}}].

\bibitem{Tremaine:1979we}
S.~Tremaine and J.E.~Gunn, \emph{{Dynamical Role of Light Neutral Leptons in
  Cosmology}}, \href{https://doi.org/10.1103/PhysRevLett.42.407}{\emph{Phys.
  Rev. Lett.} {\bfseries 42} (1979) 407}.

\bibitem{Davoudiasl:2020uig}
H.~Davoudiasl, P.B.~Denton and D.A.~McGady, \emph{{Ultralight fermionic dark
  matter}}, \href{https://doi.org/10.1103/PhysRevD.103.055014}{\emph{Phys. Rev.
  D} {\bfseries 103} (2021) 055014}
  [\href{https://arxiv.org/abs/2008.06505}{{\ttfamily 2008.06505}}].

\bibitem{Ban:2022jgm}
K.~Ban, D.Y.~Cheong, H.~Okada, H.~Otsuka, J.-C.~Park and S.C.~Park,
  \emph{{Phenomenological implications on a hidden sector from the festina
  lente bound}}, \href{https://doi.org/10.1093/ptep/ptac176}{\emph{PTEP}
  {\bfseries 2023} (2023) 013B04}
  [\href{https://arxiv.org/abs/2206.00890}{{\ttfamily 2206.00890}}].

\end{thebibliography}\endgroup

\end{document}